\RequirePackage[l2tabu]{nag}

\documentclass[a4paper,11pt,leqno]{memoir} 
\usepackage{verbatim}
\usepackage{datetime}
\usepackage{ifpdf}
\ifpdf
\fi
\ifdraftdoc 
	\usepackage{draftwatermark}				
	\SetWatermarkScale{0.3}
	\SetWatermarkText{\bf Draft: \today}
\fi
\newsubfloat{figure}
\newsubfloat{table}
\settrimmedsize{297mm}{210mm}{*}
\settypeblocksize{634pt}{448.13pt}{*} 
\setulmargins{4cm}{*}{*} 
\setlrmargins{*}{*}{1.5} 
\setmarginnotes{17pt}{51pt}{\onelineskip} 
\setheadfoot{\onelineskip}{2\onelineskip} 
\setheaderspaces{*}{2\onelineskip}{*} 
\checkandfixthelayout
\RequirePackage{etex}
\usepackage{fouriernc}
\usepackage[T1]{fontenc}
\OnehalfSpacing 
\setsecnumdepth{subsection} 
\maxsecnumdepth{subsubsection}
\usepackage{calc,soul,fourier}
\makeatletter 
\newlength\dlf@normtxtw 
\newsavebox{\feline@chapter} 
\newcommand\feline@chapter@marker[1][4cm]{%
	\sbox\feline@chapter{%
		\resizebox{!}{#1}{\fboxsep=1pt%
			\colorbox{gray}{\color{white}\thechapter}%
		}}%
		\rotatebox{90}{%
			\resizebox{%
				\heightof{\usebox{\feline@chapter}}+\depthof{\usebox{\feline@chapter}}}%
			{!}{\scshape\so\@chapapp}}\quad%
		\raisebox{\depthof{\usebox{\feline@chapter}}}{\usebox{\feline@chapter}}%
} 
\newcommand\feline@chm[1][4cm]{%
	\sbox\feline@chapter{\feline@chapter@marker[#1]}%
	\makebox[0pt][c]{
		\makebox[1cm][r]{\usebox\feline@chapter}%
	}}
\makechapterstyle{daleifmodif}{

	\renewcommand\printchapternum{\null\hfill\feline@chm[2.5cm]\par}

} 
\makeatother 
\chapterstyle{daleifmodif}
\makepagestyle{myvf} 
\makeoddfoot{myvf}{}{\thepage}{} 
\makeevenfoot{myvf}{}{\thepage}{} 
\makeheadrule{myvf}{\textwidth}{\normalrulethickness} 
\makeevenhead{myvf}{\small\textsc{\leftmark}}{}{} 
\makeoddhead{myvf}{}{}{\small\textsc{\rightmark}}
\newcommand{\clearemptydoublepage}{\newpage{\thispagestyle{empty}\cleardoublepage}}
\makeindex
\usepackage{import}

\usepackage{lipsum}					
\usepackage{amsfonts} 					
\usepackage[centertags]{amsmath}			
\usepackage{stmaryrd}					
\usepackage{amssymb}					
\usepackage{amsthm}					
\usepackage{newlfont}					
\usepackage{layouts}					
\usepackage{graphicx}					
\usepackage{longtable,rotating}			
\usepackage[applemac]{inputenc}			
\usepackage{colortbl}					
\usepackage{wasysym}					
\usepackage{mathrsfs}					
\usepackage{float}						
\usepackage{verbatim}					
\usepackage{upgreek }					
\usepackage{latexsym}					
\usepackage[square,numbers,
		     sort&compress]{natbib}		
\usepackage{url}						
\usepackage{etex}						
\usepackage[spanish,english]{babel}		
\usepackage{color}                    				
\usepackage[colorlinks=true,
		     allcolors=black]{hyperref}              
\usepackage{memhfixc}					
\usepackage{enumerate}					
\usepackage{footnote}					
\usepackage{microtype}					
\usepackage{rotfloat}					
\usepackage{alltt}						
\usepackage[version=0.96]{pgf}			
\usepackage{tikz}						
\usetikzlibrary{arrows,shapes,snakes,
		       automata,backgrounds,
		       petri,topaths}				

\usepackage{mathtools}
\DeclarePairedDelimiter\bra{\langle}{\rvert}
\DeclarePairedDelimiter\ket{\lvert}{\rangle}
\DeclarePairedDelimiterX\braket[2]{\langle}{\rangle}{#1 \delimsize\vert #2}
\newcommand{\pgftextcircled}[1]{                                                                    
    \setbox0=\hbox{#1}%
    \dimen0\wd0%
    \divide\dimen0 by 2%
    \begin{tikzpicture}[baseline=(a.base)]%
        \useasboundingbox (-\the\dimen0,0pt) rectangle (\the\dimen0,1pt);
        \node[circle,draw,outer sep=0pt,inner sep=0.1ex] (a) {#1};
    \end{tikzpicture}
}
\newcommand{\blackged}{\hfill$\blacksquare$}
\newcommand{\whiteged}{\hfill$\square$}
\newcounter{proofcount}

\let\oldsqrt\sqrt
\def\sqrt{\mathpalette\DHLhksqrt}
\def\DHLhksqrt#1#2{%
\setbox0=\hbox{$#1\oldsqrt{#2\,}$}\dimen0=\ht0
\advance\dimen0-0.2\ht0
\setbox2=\hbox{\vrule height\ht0 depth -\dimen0}%
{\box0\lower0.4pt\box2}}
\newcommand{\mycaption}[2][\@empty]{
	\captionnamefont{\scshape} 
	\changecaptionwidth
	\captionwidth{0.9\linewidth}
	\captiondelim{.\:} 
	\indentcaption{0.75cm}
	\captionstyle[\centering]{}
	\setlength{\belowcaptionskip}{10pt}
	\ifx \@empty#1 \caption{#2}\else \caption[#1]{#2}
}
\newcommand{\mysubcaption}[2][\@empty]{
	\subcaptionsize{\small}
	\hangsubcaption
	\subcaptionlabelfont{\rmfamily}
	\sidecapstyle{\raggedright}
	\setlength{\belowcaptionskip}{10pt}
	\ifx \@empty#1 \subcaption{#2}\else \subcaption[#1]{#2}
}
\usepackage{lettrine}
\newcommand{\initial}[1]{%
	\lettrine[lines=3,lhang=0.33,nindent=0em]{
		\color{gray}
     		{\textsc{#1}}}{}}
\theoremstyle{plain}

\theoremstyle{plain}

\theoremstyle{plain}
\theoremstyle{definition}

\theoremstyle{plain}

\theoremstyle{plain}

\theoremstyle{plain}

\usepackage{hyperref}
\usepackage{siunitx}
\usepackage{natbib}
\usepackage[english]{babel}
\usepackage{siunitx}

\hypersetup{
    colorlinks=true,
    linkcolor=blue,
    filecolor=magenta,      
    urlcolor=cyan
}

\begin{document}

\frontmatter
\pagenumbering{roman}

\begin{titlingpage}
\begin{SingleSpace}
\calccentering{\unitlength} 
\begin{adjustwidth*}{\unitlength}{-\unitlength}
\vspace*{1.5mm}
\begin{center}
\rule[0.5ex]{\linewidth}{2pt}\vspace*{-\baselineskip}\vspace*{3.2pt}
\rule[0.5ex]{\linewidth}{1pt}\\[\baselineskip]
{\HUGE Tunneling in Graphene SymFETs \\[4mm]}
\rule[0.5ex]{\linewidth}{1pt}\vspace*{-\baselineskip}\vspace{3.2pt}
\rule[0.5ex]{\linewidth}{2pt}\\
\vspace{5mm}
{\large By}\\
\vspace{8mm}
{\large\textsc{\textbf{Supratik Sarkar}}}\\
{\large Class Roll No. 001410701066}\\
{\large Exam Roll No. ETC188042}\\
{\large Registration No. 127717 of 2014-2015}\\
\vspace{5mm}
\large AND \\
\vspace{4.5mm}
{\large\textsc{\textbf{Samrat Sarkar}}}\\
{\large Class Roll No. 001410701064}\\
{\large Exam Roll No. ETC188061}\\
{\large Registration No. 127715 of 2014-2015}\\
\vspace{8mm}
{\textsc{Under the supervision of}}\\
{\large \textsc{\textbf{Prof. Chayanika Bose}}}\\
\vspace{10mm}

{\textsc{Electronics and Telecommunication Engineering}\\
\textsc{Jadavpur University}}\\
\textsc{Kolkata-700032}\\
\vspace{11mm}
\begin{minipage}{14cm}
\begin{center}
\textsc{Project report submitted in partial fulfillment of the requirements for the degree of Bachelor of Enginering in Electronics and Telecommunication Engineering }
\end{center}
\end{minipage}\\
\vspace{18mm}
{\large \textsc{MAY 2018}}
\vspace{12mm}
\end{center}
\end{adjustwidth*}
\end{SingleSpace}
\end{titlingpage}
\clearemptydoublepage
%
%
%
%
%
%
\chapter*{Certificate}
\begin{SingleSpace}
\begin{quote}
\initial{T}his is to certify that the project report entitled \textbf{"Tunneling in Graphene SymFETs"} submitted by \textbf{Supratik Sarkar}, Exam Roll No. \textbf{ETC188042} and \textbf{Samrat Sarkar}, Exam Roll No. \textbf{ETC188061} for the partial fulfillment of the degree of Bachelor of Electronics and Telecommunication Engineering of Jadavpur University is based on the assigned Project Work during the session \textbf{2017-2018} under the supervision of \textbf{Prof. Chayanika Bose} of \textbf{Electronics and Telecommunication Engineering Department} of \textbf{Jadavpur University}, Kolkata, West Bengal, India.

\vspace{1.3cm}

\noindent 
\end{quote}
\end{SingleSpace}
\clearpage
\clearemptydoublepage
%
%

\chapter*{Acknowledgements}
\begin{SingleSpace}
\initial{W}e would like to extend our gratitude to a lot of people who have directly or indirectly contributed to the completion of this project. First and foremost, we would like to thank our supervisor Prof. Chayanika Bose for her constant encouragement and support. This would not have been possible without her guidance. We are also indebted to the entire faculty of the Department of Electronics and Telecommunication Engineering, Jadavpur University, for providing us with the required infrastructure which acted as the backbone of this project. Last but not the least, we would like to thank our families, friends and well-wishers whose support was crucial towards the completion of this project.
\end{SingleSpace}
\clearpage
\clearemptydoublepage
\chapter*{Abstract}
\begin{SingleSpace}
\initial{W}ith the further scaling of silicon MOSFETs becoming increasingly harder, the search for an alternative material became crucial. The electron device community found many of the answers in two dimensional materials, especially graphene. With an astounding mobility and perfectly symmetrical bandstructure, graphene may be, just the replacement for silicon we have been looking for. In this report, the mechanism of tunneling in a graphene-insulator-graphene (GIG) junction has been studied, by applying Bardeen's transfer Hamiltonian approach. Later, the formalism of the GIG junction has been used to study the performance and current-voltage characteristics of a symmetric tunneling field effect transistor or SymFET. The device exhibits a small tunneling current at most of the biasing voltages. But when the Dirac points of the oppositely doped graphene layers are aligned, a large amount of tunneling current is observed. The performance of the device has been studied for various device dimensions. The resonant current peak is also shown to increase for higher levels of doping. The extraordinary symmetry of the $I-V$ characteristics makes SymFET a potential candidate for high speed analog devices. The SymFET is also shown to be robust to temperature changes, since tunneling is the main mechanism of charge transport. With further study and modifications, the SymFET can become a popular choice for both analog and digital circuit implementation. 
\end{SingleSpace}
\clearpage
\clearemptydoublepage

\renewcommand{\contentsname}{Table of Contents}
\maxtocdepth{section}
\tableofcontents*
\addtocontents{toc}{\par\nobreak \mbox{}\hfill{\bf Page}\par\nobreak}
\clearemptydoublepage
\listoffigures
\addtocontents{lof}{\par\nobreak\textbf{{\scshape Figure} \hfill Page}\par\nobreak}
\clearemptydoublepage


\mainmatter
%
%

\let\textcircled=\pgftextcircled
\chapter{Introduction}
\label{chap:intro}

\initial{G}raphene is a thin two dimensional sheet of crystalline carbon, only a few atoms thick and was first reported by Novoselov \textit{et al.} in 2004. It shows novel properties like very high mobility, exceeding those of commonly used semiconductor materials, and perfect two dimensional confinement. This has earned graphene the mantle of a "supermaterial" and it is often considered as the best candidate for the post-silicon electronic technologies.

One of the most fascinating properties of graphene is its perfectly symmetric bandstructure with the valence and conductance bands mirroring each other. The conical shaped valence band minimas touch the conduction band at Dirac point. In undoped graphene, the Fermi energy lies exactly on the Dirac points and the Fermi surface consists of the Dirac points. Thus, graphene can be called a zero-gap semi-metal (since it has sufficient conductivity even in ultracold regime) and opening a finite bandgap in graphene is a big challenge.

Until 2009, most of the research on graphene exploited only the 2D nature of the newly proposed supermaterial. In 2009, Banerjee \textit{et al.} proposed out of plane charge transport  in a bilayer pseudo-spin FET (BiSFET). BiSFET exhibited the property of charge conduction between two graphene monolayers, by varying the tunnel resistance between them by the property of excitonic condensation. The excitonic condensation of graphene can also occur at room temperature. If one layer of graphene populated with electrons is brought in close proximity to another layer of graphene populated with holes, then the Coulomb interaction between the layers causes excitonic condensation. This leads to a significant tunneling current between the monolayers. The BiSFETs are considered as superior candidates for digital logic circuits because of their low power dissipation.

In this report, we will theoretically examine the properties of tunneling across a graphene-insulator-graphene (GIG) tunnel junction. We will examine the case, when one of the sheets is doped n-type and the other is doped p-type. Then we shall study the current voltage relationship in such devices and extrapolate them to the formation of a symmetric tunneling FET (SymFET). The symmetric resonant peaks of the SymFET make it an extremely good candidate for high speed analog electronics and can be used to implement digital logic, just like BiSFET. They are also very fast (since they conduct by tunneling) and are pretty robust to temperature fluctuations, but have relatively poor ON-OFF ratio. The device exhibits a small tunneling current at most of the biasing values, when energy and momentum are conserved only for a single energy level, lying midway between the Fermi levels. But, if the Dirac points of the graphene sheets are properly aligned, a very large amount of tunneling current flows through the device, as momentum and energy are conserved for all energy values between the Fermi levels of the two layers of graphene. 

But before we delve into the details about tunneling in a typical GIG structure and SymFET, we will like to add a small prelude on some of the important properties of graphene and why it is considered to be a rising star among other materials, in the electronics manufacturing industry.

\section{Present day scenario of electronic devices}

Till date, metal oxide semiconductor field effect transistor (MOSFET) is the most used and versatile electronic device used for both digital and radio frequency (RF) applications. To achieve the goal of higher functionality and more powerful devices, we have been on the quest of miniaturizing our transistors and increasing the degree of circuit integration for the past many decades. The trend of doubling the capacity of circuit integration every $1.5-2$ years has continued for the past $50$ years, following the predictions given by Gordon Moore, the co-founder of Intel. However. since approximately $2010$, it has become evident that scaling of MOSFET devices will no longer be possible, beyond a certain limit. As we decrease the channel length to get faster devices and achieve higher packing density, non-linear effects start creeping into the MOSFET devices and leakage power increases. Since the scaling is likely to reach its fundamental limit in not so far a future, an alternative to silicon (Si) technology has become a global demand. 

\section{Exciting properties of graphene}

While the electron-devices community were fighting to combat the effects of non-linear short channel effects, it was noticed that two dimensional (2D) materials don't show short channel non-linear effects even if the transport channel length is too small. Therefore, these materials are a good choice as an alternative of Si. Graphene is a very interesting 2D material which possesses many beneficial properties.

The physical structure of graphene is like a honeycomb lattice. The structure is completely planar, i.e., 2D. The distance between carbon atoms is $1.42$ \si{\angstrom}. The three $sp^2$ hybridized orbitals, which are symmetrically distributed (at angles of $120$ degree), form three sigma-bonds with those of the three nearest carbon atoms. The strength of the sigma-bonds makes graphene one of the strongest materials. The orbitals of the remaining $p_z$ electrons are distributed perpendicular to the molecular plane and they form what is known as the pi-bonds with those of one of the three nearest carbon atoms. These bonds can have two different orientations and hence the graphene structure can be viewed as two interpenetrating triangular sub lattices $A$ and $B$ with two atoms per unit cell.

\begin{figure}[H]
	\begin{center}
		\includegraphics[width=0.7\textwidth]{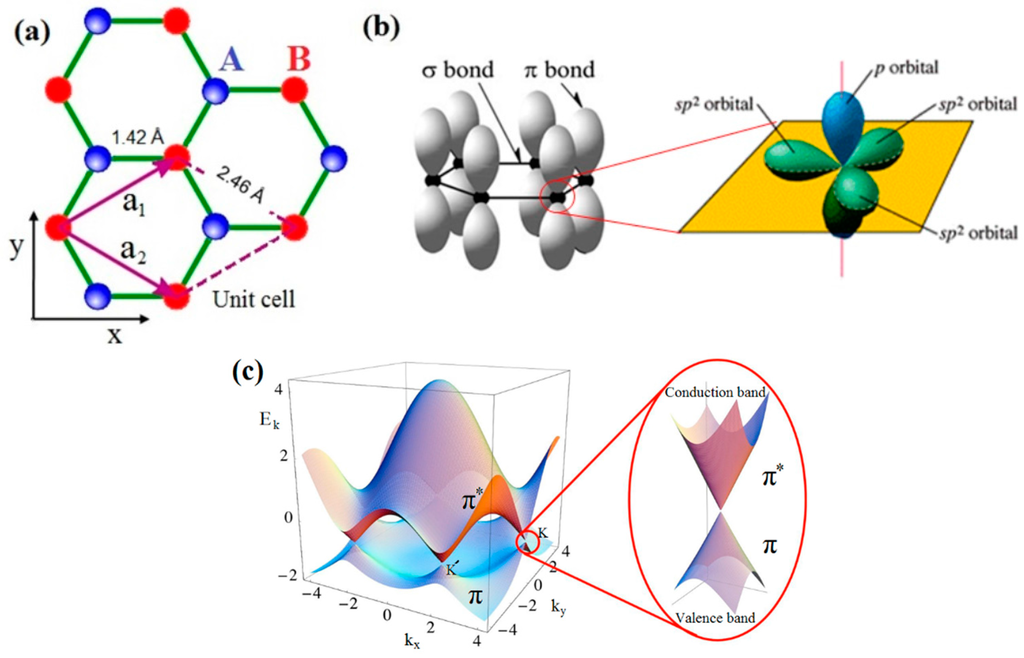}
		\mycaption[Lattice structure and chemical bonds in graphene]{(a) Bravais lattice of the graphene, (b) $\sigma$ and $\pi$ bonds in graphene and (c) graphene $\pi-$ and $\pi^*-$band structure.}
	\end{center}
\end{figure}

Graphene has zero band gap due to inversion symmetry in its physical planar structure. The meeting points of the valence and conduction bands are known as the Dirac point. Graphene has a linear dispersion relation, at least near the Dirac points. Such dispersion relation is an attribute of zero mass particles like the photons. It is thus reminiscent of the fact that the electrons in graphene behave like massless Dirac fermions moving with an effective velocity of light equal to the Fermi velocity. The energy band is exactly symmetric about the null kinetic energy point , and this condition is met only at the two Dirac points, it follows that for exactly half filling of the band the density of state at the Fermi level is exactly zero. But in the absence of doping graphene has exactly one electron per "spin" per atom (2 per unit cell), so taking spin into account the band is indeed exactly half filled. Thus, undoped graphene is a perfect semimetal!

In contrast to the step-like density of states exhibited by 2D electron gas, graphene shows a linear density of states, although it is a 2D material. Due to the massless nature and velocity comparable to that of light ($0.3\%$ the velocity of light in vacuum), electron motion in graphene is governed by the relativistic Dirac like equation instead of the non-relativistic Schr\"odinger equation. While Dirac equation has a four-component spinor wave function that accounts for the two spin states of electron (particle) and positron (antiparticle), the Dirac-like equation for graphene has two-component pseudo-spinor wave function that determines the relative electron population on the two lattice sites A and B. The spinor in the latter case has nothing to do with electron spin and hence is usually referred to as the pseudo spinor state.

\begin{figure}[]
	\begin{center}
		\includegraphics[width=0.8\textwidth]{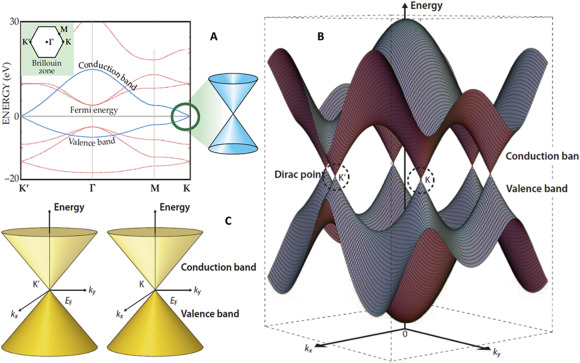}
		\mycaption[Band structure near the Fermi level of graphene. (A) 2D schematic diagram, (B) 3D schematic diagram and (C) Dirac cone of K and K', which correspond to the Fermi level of (B)]{Band structure near the Fermi level of graphene. (A) 2D schematic diagram, (B) 3D schematic diagram and (C) Dirac cone of K and K', which correspond to the Fermi level of (B).}
	\end{center}
\end{figure}

One of the distinctive features of graphene is the absence of back scattering as in carbon nanotube. It turns out that the carrier mobility in graphene is astoundingly high. Measured values show that it could be as high as $2.3\times 10^5$ cm$^2$V$^{-1}$s$^{-1}$. But such high mobility is possible only for pristine samples where care has been taken to minimize the scattering centers, so that carrier transport is entirely ballistic. In most of the practical samples however, the carrier transport is strongly limited by scattering from extrinsic sources, thus making the transport diffusive. The charged impurities on the graphene surface as well as in the graphene-substrate interface, and the interfacial and substrate phonons constitute the external sources of scattering. These scattering centers limit the carrier mobility drastically to values ranging from $10000$ cm$^2$V$^{-1}$s$^{-1}$ (CVD graphene transferred to SiO and epitaxial graphene on SiC) to $15000$ cm$^2$V$^{-1}$s$^{-1}$ (exfoliated graphene on SiO) under ambient condition.

Carrier velocity is also an important transport parameter. For low source to drain field, the velocity first linearly increases with increase in electric field as long as the transport is dominated by the elastic scattering processes discussed above. But at higher electric field the inelastic optical phonon scattering increases and the velocity saturates. Nonetheless, under the most ideal condition the carrier velocity in graphene may approach as high as the Fermi velocity ($10^6$ ms$^{-1}$). The high mobility and high velocity in graphene are particularly attractive for high frequency circuits (both digital and analog). Thus in principle, graphene should be a potential candidate for the FET industry.

Concomitant to the large mobility, graphene conductivity is also high and is observed to be better than that of Cu. With a most conservative estimate, the conductivity of graphene, at a carrier density of $10^{12}$ m$^{-2}$ is arrived at $9.6\times 10^5$ \si{\ohm}$^{-1}$cm$^{-1}$ as against a value of $6\times 10^5$ \si{\ohm}$^{-1}$cm$^{-1}$ for Cu. But the more interesting part is the conductivity modulation with gate bias. Surprisingly the conductivity never goes to zero even at zero gate bias when the carrier density goes to zero. It is now well understood that graphene transport near the Dirac points at low carrier density is dominated by a random distribution of carrier inhomogeneity referred to as "electron and hole puddles". Even at the zero gate voltage, these puddles cannot vanish all at a time. Therefore, although the average carrier density goes to zero, some electron and hole puddles still remain. As a result, the conductivity never goes to zero. 

It is edifying to note that the same graphene sample could be p-type or n-type depending on the gate bias applied. This implies that the carriers could be continuously tuned from electron to hole and vice versa by adjusting the applied gate bias. Such electric field induced carrier type as well as carrier density modulation shall offer a great deal of flexibility to the device designers to conceive a p-n junction, without physically doping the material.

\section{Graphene FETs}

With the discovery of graphene, researchers were very hopeful that it would be a potential candidate to replace silicon. But very soon graphene tumbled into difficulty. Being gapless, graphene does not allow the FET to switch off resulting in a high leakage current and prohibitive energy dissipation. Several attempts to induce bandgap in graphene include cutting the graphene into nanoribbons, surface functionalization, subjecting bi-layer graphene to electric field, etc. These attempts nonetheless, have resulted in bandgaps of few hundred meV only whereas practically to make graphene suitable for digital logic devices, it requires a bandgap on the order of an eV at room temperature. Most researchers tried to change the bandgap in graphene to make it more suitable for application in logic circuits, but the outcome of these efforts resulted in degradation of graphene properties like the mobility, which was the prime attraction. Although a density functional theory calculation predicts that a bandgap of 1.2 eV can be induced in graphene through surface functionalization, the experimental observation is contrary to the prediction. The absence of required bandgap makes it difficult to achieve suitable on-off switching ratio for low power dissipation.

Alternative graphene transistor architecture by Britnell \textit{et al.} based on quantum tunneling from a graphene electrode through thin insulating barrier layers of hexagonal boron nitride (hBN) and molybdenum disulfide, reported room-temperature high switching ratios. Such graphene FETs have shown potential for high-frequency operation and large-scale integration. The switching ratio can be enhanced with optimized architecture and has opened a new area of research to explore the prospects of field effect tunneling transistors for possible applications in graphene nanoelectronics. 

At present one of the most influential graphene FET architectures that provide high on-off switching ratio was proposed by Zhao \textit{et al.} and is known as the Graphene Symmetric FET or SymFET. It is a vertical tunnel field effect transistor which uses an insulator layer sandwiched between two graphene monolayers. The device works on the principle of GIG tunneling. We will discuss the theory of GIG structure and SymFET in details in the next chapters. 

%
%

\let\textcircled=\pgftextcircled
\chapter{Mathematical Formalisms}
\label{chap:formalism}

\section{Bardeen transfer Hamiltonian}

\begin{figure}[H]
	\begin{center}
		\includegraphics[width=0.4\textwidth]{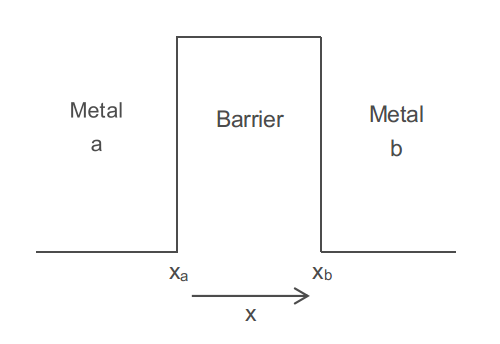}
		\mycaption[Energy profile of a simple metal-barrier-metal system]{Energy profile of a simple metal-barrier-metal system.}
		\label{fig:1}
	\end{center}
\end{figure}

\initial{I}n this section we will review the Bardeen transfer Hamiltonian approach (the same Bardeen who was awarded the Nobel Prize for the invention of transistors) given by J. Bardeen in 1960. Let us consider the system shown in Fig. \ref{fig:1}. Here the barrier extends from $x_a$ to $x_b$ along the $x$ axis. There is metal $a$ to the left of $x_a$ and metal $b$ to the right of $x_b$. We can consider two many-particle states of the entire metal-barrier-metal system: $\psi_0$ and $\psi_{mn}$. $\psi_{mn}$ differs from $\psi_0$ in the transfer of an electron from state $m$ in metal $a$ to state $n$ in metal $b$. This leaves a hole in $m$ in $\psi_{mn}$. We must note that, the states $\psi_0$ and $\psi_{mn}$ can be specified by their quasi-particle occupation numbers in $a$ and $b$.

The quasi-particles do not correspond to plane waves, but to waves which are reflected at the barrier and attenuate exponentially inside the barrier. By WKB approximation, the wavefunction can be expressed as:

\begin{subequations}
   \begin{equation}
   \label{eq1}
   \psi_m=Cp_x^{-1/2}e^{i\left(p_yy+p_zz\right)}\sin{\left(p_xx+\phi\right)} \quad \text{for $x<x_a$ (in metal $a$)}
   \end{equation}
   \begin{equation}
   \label{eq2}
   \psi_m=\dfrac{1}{2}C|p_x|^{-1/2}e^{i\left(p_yy+p_zz\right)}\times\exp\left(-\int_{x_a}^{x}|p_x|dx\right) \quad \text{for $x_a<x<x_b$ (in barrier)} 
   \end{equation}
\end{subequations}

\noindent where $C=\sqrt{\frac{2p_x}{L}}$ is the normalization constant and in the barrier region $\lvert p_x \rvert = \sqrt{2\mu U -p_y^2 -p_z^2}$, where $U\left(x\right)$ is the potential energy. To get a good solution for $x>x_b$, we assume that $\psi_m$ smoothly drops to zero beyond $x_b$.

Thus, let $\psi_0$ be the solution of the Schr\"odinger equation with energy $W_0$ to the left of $x=x_b$. Similarly, let $\psi_{mn}$ with energy $W_{mn}$ be the solution to the left of $x_a$, where wavefunction $\psi_n$ smoothly drops to zero. However, in the barrier both $\psi_0$ and $\psi_{mn}$ are applicable. So, the time dependent solution can be expressed as the linear combination of both the wavefunctions $\psi_0$ and $\psi_{mn}$:

\begin{equation}
\label{eq3}
\psi=a\left(t\right)\psi_0e^{-iW_0t}+\sum_{m,n}b_{mn}\left(t\right)\psi_{mn}e^{-iW_{mn}t}
\end{equation}  

If we solve this wavefunction $\psi$, by substituting it in the Schr\"odinger equation, we get the transformation matrix elements as:

\begin{equation}
\label{eq4}
M_{mn}=\int\psi_0^*\left(H-W_{mn}\right)\psi_{mn}d\tau
\end{equation} 

\noindent Since, $\psi_{mn}$ is the solution for $x<x_a$, the integrand becomes zero to the right of $x_a$ and it reduces to:

\begin{equation}
\label{eq5}
M_{mn}=\int_{x<x_a}\psi_0^*\left(H-W_{mn}\right)\psi_{mn}d\tau
\end{equation}

\noindent Similarly, we can note that the expression $\psi_{mn}\left(H-W_{0}\right)\psi_0^*$ is zero to the left of $x_b$. So, subtracting it from the integrand of Eq. \ref{eq5} is a valid operation and it gives us:

\begin{equation}
\label{eq6}
M_{mn}=\int_{x<x_a}\left\lbrace\left(\psi_0^*H\psi_{mn}-\psi_{mn}H\psi_0^*\right)+\left(\psi_{mn}W_0\psi_0^*-\psi_0^*W_{mn}\psi_{mn}\right)\right\rbrace d\tau
\end{equation}

\noindent We are interested in the final states, where $W_{mn}\approx W_0$. S, the transformation matrix element reduces to the more symmetric form, as given by:

\begin{equation}
\label{eq7}
M_{mn}=\int_{x<x_a}\left(\psi_0^*H\psi_{mn}-\psi_{mn}H\psi_0^*\right)d\tau
\end{equation}

\noindent The transition probability of electron from one side to the other is given by the expression $\left(2\pi/\hbar\right)|M|^2\rho_f$ where $M$ is the matrix element and $\rho_f$ is the energy density of final states.

\section{Graphene insulator graphene junctions}

In this section, we shall deal with the formalisms of tunneling in a graphene insulator graphene (GIG) junction. Let us assume that the left hand electrode is n-doped and the electrode on the right is p-doped. The Fermi level of the left hand electrode is given by $\mu_L=E_{DL}+\Delta E_L$, where the subscript $L$ denotes the properties of left hand electrode. Similarly, the Fermi level of the right hand electrode is given by $\mu_R=E_{DR}-\Delta E_R$. $E_{DL}$ and $E_{DR}$ are the Dirac points of the respective electrodes. If we apply a bias voltage $V$ between the electrodes, we get $\mu_L-\mu_R=eV$. For simplicity, let us assume $\Delta E_L=\Delta E_R=\Delta E>0$. 

\begin{figure}[H]
	\begin{center}
		\includegraphics[width=0.7\textwidth]{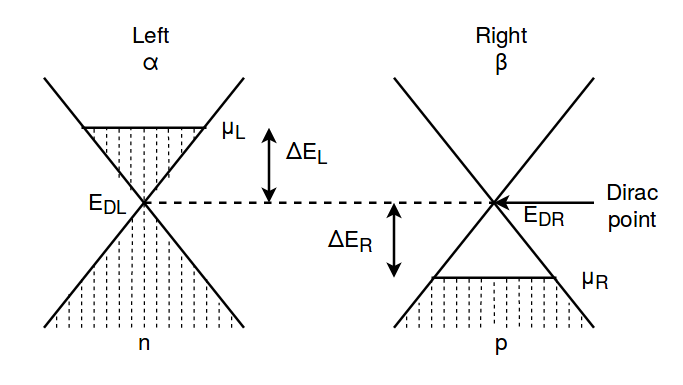}
		\mycaption[Band diagram for a doped GIG junction where the Dirac points are in perfect alignment]{Band diagram for a doped GIG junction where the Dirac points are in perfect alignment.}
		\label{fig:2}
	\end{center}
\end{figure}

If $eV\neq 2\Delta E$, then, momentum is conserved only for a single energy level, midway between the Dirac points of the electrodes. Intuitively, we can see that the circumference of the level at which momentum is conserved varies linearly with the voltage applied externally, and this gives a linear $I-V$ characteristic. On the contrary, if $eV=2\Delta E$, then \textbf{k} conservation holds across all energy levels and a large amount of current flows between the electrodes. This resonant state is the condition we are most interested in.

\subsection{Tunneling formalism}

Using Bardeen's approach, the tunneling current is given upto first order as (detailed proof in Appendix \ref{app:app01}):

\begin{equation}
\label{eq8}
I=g_Sg_Ve\sum_{\alpha,\beta}\left\lbrace\dfrac{1}{\tau_{\alpha\beta}}f_L\left(E_{\alpha}\right)\left[1-f_R\left(E_{\beta}\right)\right]-\dfrac{1}{\tau_{\beta \alpha}}f_R\left(E_{\beta}\right)\left[1-f_L\left(E_{\alpha}\right)\right]\right\rbrace
\end{equation} 

\noindent where $\alpha$ and $\beta$ stand for states in left and right electrodes simultaneously, $E_{\alpha}$ and $E_{\beta}$ are the energies of the electrodes, $g_S=2$ is the spin degeneracy, $g_V$ is the valley degeneracy, $1/\tau_{\alpha,\beta}$ and $1/\tau_{\alpha,\beta}$ are the tunneling rates from left to right and right to left simultaneously, and $f_L$ and $f_R$ are the Fermi factors for the electrodes, where 

\begin{equation}
\label{eq9}
f_{L/R}\left(E\right)=\dfrac{1}{1+\exp\left[\left(E-\mu_{L/R}\right)/k_BT\right]}
\end{equation}

\noindent The tunneling rates $1/\tau_{\alpha,\beta}$ and $1/\tau_{\alpha,\beta}$ are equal as:

\begin{equation}
\label{eq10}
\dfrac{1}{\tau_{\alpha\beta}}=\dfrac{2\pi}{\hbar}|M_{\alpha\beta}|^2\delta\left(E_{\alpha}-E_{\beta}\right)=\dfrac{1}{\tau_{\beta\alpha}}
\end{equation}

\noindent where 

\begin{equation}
\label{eq11}
M_{\alpha\beta}=\dfrac{\hbar^2}{2m}\int \left(\psi_{\alpha}^*\frac{d\psi_{\beta}}{dz}-\psi_{\beta}\frac{d\psi_{\alpha}^*}{dz}\right)dS
\end{equation}

\noindent which we directly get from Eq. \ref{eq7}, is the transition matrix element, where $\psi_{\alpha}\left(\textbf{r},z\right)$ and $\psi_{\beta}\left(\textbf{r},z\right)$ are the wavefunctions of the left and right electrodes.

Now, for the next step, let us consider two graphene atoms in a unit cell. The wavevector can be expressed as the superposition of the orthogonal basis elements $\phi_{j\textbf{k}}$ of each atom, such that:

\begin{equation}
\label{eq12}
\psi\left(\textbf{r},z\right)=\chi_1\left(\textbf{k}\right)\phi_{1\textbf{k}}\left(\textbf{r},z\right)+\chi_2\left(\textbf{k}\right)\phi_{2\textbf{k}}\left(\textbf{r},z\right)
\end{equation}  

\noindent If $A$ is the area of the electrode, then $\phi_{j\textbf{k}}\left(\textbf{r},z\right)=\exp\left(i\textbf{k}\cdot\textbf{r}\right)u_{j\textbf{k}}\left(\textbf{r},z\right)/\sqrt{A}$, where $u_{j\textbf{k}}\left(\textbf{r},z\right)$ is a periodic function. Thus, the states in the left electrode can be expressed as:

\begin{equation}
\label{eq13}
\begin{split}
\psi_{\alpha}&=\chi_{1,\alpha}\left(\textbf{k}\right)\phi_{1\textbf{k},\alpha}\left(\textbf{r},z\right)+\chi_{2,\alpha}\left(\textbf{k}\alpha\right)\phi_{2\textbf{k},\alpha}\left(\textbf{r},z\right) \\
&= \dfrac{1}{\sqrt{A}}e^{i\textbf{k}_{\alpha}\cdot\textbf{r}}\left[  \chi_{1,\alpha}\left(\textbf{k}\right)u_{1\textbf{k},\alpha}\left(\textbf{r},z\right)+\chi_{2,\alpha}\left(\textbf{k}\right)u_{2\textbf{k},\alpha}\left(\textbf{r},z\right)\right]
\end{split}
\end{equation}

\noindent Using Eq. \ref{eq13}, the expression involving $u_{1\textbf{k},\alpha}\left(\textbf{r},z\right)$ of $\psi_{\alpha}\left(\textbf{r},z\right)$ and $u_{1\textbf{k},\alpha}\left(\textbf{r},z\right)$ of $\psi_{\beta}\left(\textbf{r},z\right)$ in $M_{\alpha\beta}$, as given in Eq. \ref{eq11}, we get:

\begin{equation}
\label{eq14}
\begin{split}
&\int e^{-i\textbf{k}_{\alpha}\cdot\textbf{r}}\times e^{i\textbf{k}_{\beta}\cdot\textbf{r}}\left[ u_{1\textbf{k},\alpha}^*\frac{du_{1\textbf{k},\beta}}{dz}-u_{1\textbf{k},\beta}\frac{du_{1\textbf{k},\alpha}^*}{dz} \right] dS \\
& \approx \dfrac{2\kappa e^{-\kappa d}}{D} u_{11}^2 \int e^{i\left(\textbf{k}_{\beta}-\textbf{k}_{\alpha}\right)\cdot\textbf{r}}dS
\end{split}
\end{equation}

\noindent where $u_{11}^2$ is a constant of order unity and $2\kappa e^{-\kappa d}/D$ gives the regular tunneling expression in the $z$ direction, where $d$ is the separation between the electrodes, $\kappa$ is the decay constant of the wavefunctions in the barrier/insulator and $D$ is a normalization constant. The expression $u_{j\textbf{k}}\left(\textbf{r},z\right)$ is a very weak function of the radial parameter $\textbf{r}$, the radial dependence has been approximated into numerical vales, to get to the expression in Eq. \ref{eq14}.

\begin{figure}[H]
	\begin{center}
		\includegraphics[width=0.55\textwidth]{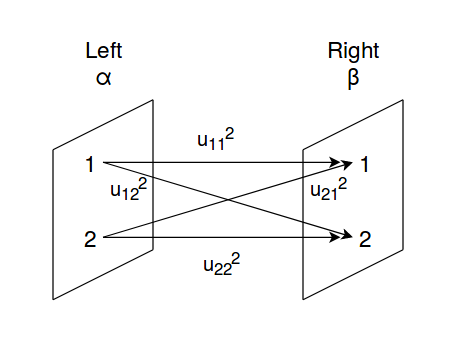}
		\mycaption[Visual depiction of the constants $u_{ij}$ for $i\in\left\lbrace1,2\right\rbrace$ and $j\in\left\lbrace1,2\right\rbrace$]{Visual depiction of the constants $u_{ij}$ for $i\in\left\lbrace1,2\right\rbrace$ and $j\in\left\lbrace1,2\right\rbrace$.}
		\label{fig:3}
	\end{center}
\end{figure}

Similarly, we will get constants $u_{22}$, $u_{12}$ and $u_{21}$ for the rest of the three cases. By symmetry of the underlying structure we conclude that $u_{11}=u_{22}$. Also, the constants received from the cross terms will be equal, i.e., $u_{12}=u_{21}$. The values of $\chi_1\left(\textbf{k}\right)$ and $\chi_2\left(\textbf{k}\right)$ for graphene, using nearest-neighbour tightbinding approximation, are given by:

\begin{equation}
\label{eq15}
	\begin{bmatrix}
	\chi_1 \\ \chi_2
	\end{bmatrix}
	=
	\dfrac{1}{\sqrt{2}}
	\begin{bmatrix}
	e^{\mp i\theta_{\text{k}}/2} \\ se^{\pm i\theta_{\text{k}}/2}
	\end{bmatrix}
\end{equation}

\noindent where $\theta_{\textbf{k}}$ is the angle of the relative wavevector, $s=+1$ for conduction band and $s=-1$ for valence band. For, rotationally misaligned electrodes, we find the matrix elements to be

\begin{equation}
\label{eq16}
	M_{\alpha\beta}=\dfrac{\hbar^2\kappa}{2AmD}e^{-\kappa d} g_{\omega}\left(\theta_L-\theta_R\right) \int dS e^{i\textbf{Q}\cdot\textbf{r}}e^{i\left(\textbf{k}_{\beta}-\textbf{k}_{\alpha}\right)\cdot\textbf{r}}
\end{equation}

\noindent where 

\begin{equation}
\label{eq17}
\begin{split}
g_{\omega}\left(\theta_L-\theta_R\right)&=u_{11}^2\left(e^{i\left(\theta_L \mp \theta_R'\right)/2}+s_Ls_Re^{-i\left(\theta_L \mp \theta_R'\right)/2}\right)\\&+u_{12}^2\left(s_Re^{i\left(\theta_L \pm \theta_R'\right)/2}+s_Le^{-i\left(\theta_L \mp \theta_R'\right)/2}\right)
\end{split}
\end{equation}

\noindent where $\theta_R'=\theta_R+\omega$ and  $\theta_L'=\theta_L+\omega$, for $\omega=2\sin^{-1}\left(3aQ/8\pi\right)$. $\textbf{Q}$ is the vector difference between the extremas of respective electrodes. For non-zero value of $\textbf{Q}$ the values of $u_{ij}$ constants will change, but that will not cause any significant difference. If proper alignment is done, then $\textbf{Q}=0$. Considering proper alignment for simplicity, we get:

\begin{equation}
\label{eq18}
M_{\alpha\beta}=\dfrac{\hbar^2\kappa}{2AmD}e^{-\kappa d} g_{0}\left(\theta_L-\theta_R\right) \int dS e^{i\left(\textbf{k}_{\beta}-\textbf{k}_{\alpha}\right)\cdot\textbf{r}}
\end{equation}

\noindent It is interesting to note that in the limiting case of $A\rightarrow\infty$ the integral becomes the delta function $\delta\left(\textbf{k}_{\alpha}-\textbf{k}_{\beta}\right)$. Let us define the quantity:

\begin{equation}
\label{eq19}
\Lambda\left(\Delta\textbf{k}\right)\equiv\big\lvert\frac{1}{A}\int dSe^{i\Delta\textbf{k}\cdot\textbf{r}}\big\rvert^2
\end{equation}

\noindent where $\Delta\textbf{k}=\textbf{k}_{\beta}-\textbf{k}_{\alpha}$. For $A\rightarrow\infty$, $\Lambda\left(\Delta\textbf{k}\right)$ becomes a delta function.

\begin{figure}[H]
	\begin{center}
		\includegraphics[width=0.7\textwidth]{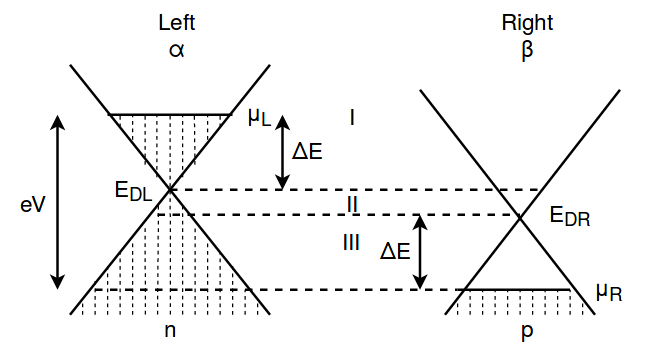}
		\mycaption[Band diagram of a doped GIG junction, highlighting the different zones of band alignments I, II and III.]{Band diagram of a doped GIG junction, highlighting the different zones of band alignments I, II and III. Here $eV>2\Delta E.$}
		\label{fig:4}
	\end{center}
\end{figure}

Setting, the value of valley degeneracy equal to $2$ for graphene and substituting the value of the tunneling rates and $M_{\alpha\beta}$ in Eq. \ref{eq8} we get:

\begin{equation}
\label{eq20}
I=\dfrac{8\pi e}{\hbar} \left(\dfrac{\hbar^2\kappa}{2mD}e^{-\kappa d}\right)^2 \sum_{B}\sum_{\textbf{k}_{\alpha},\textbf{k}_{\beta}}\lvert g_0 \left(\theta_L,\theta_R\right)\rvert^2\left[f_L\left(E_{\textbf{k},{\alpha}}\right)-f_R\left(E_{\textbf{k},{\beta}}\right)\right]\delta\left(E_{\textbf{k},{\alpha}}-E_{\textbf{k},{\beta}}\right)\Lambda\left(\Delta\textbf{k}\right)
\end{equation}

\noindent here the summation over $B$ represent the different zones of the band alignments, marked I, II and III in Fig. \ref{fig:4}.

From the figure we can write down, in zone I, $E_{\textbf{k},\alpha}=E_{DL}+\hbar v_Fk_{\alpha}$ and $E_{\textbf{k},\beta}=E_{DR}+\hbar v_Fk_{\beta}$, where $v_F$ is the Fermi velocity. So, $E_{\textbf{k},\alpha}-E_{\textbf{k},\beta}=E_{DL}-E_{DR}+\hbar v_F\left(k_{\alpha}-k_{\beta}\right)=eV-2\Delta E+\hbar v_F\left(k_{\alpha}-k_{\beta}\right)$. We define $V'=eV-2\Delta E$, such that $k_{\beta}=k_{\alpha}+eV'/\hbar v_F$. We get the same results in zone III. However in zone II, $k_{\beta}=e|V'|/\hbar v_F - k_{\alpha}$.  In the limit of large area, $\Lambda\left(\Delta\textbf{k}\right)$ becomes a delta function and $|g_{0} \left(\theta_L,\theta_R\right)|$ is replaced by $|g_{0} \left(\theta_{\textbf{k}},\theta_{\textbf{k}}\right)|=2u_{12}^2i\sin\theta_{\textbf{k}}$.

\subsection{Tunneling current for large graphene sheets}

For $A\rightarrow\infty$, $\Lambda\left(\Delta\textbf{k}\right)=\delta\left(\textbf{k}_{\beta}-\textbf{k}_{\alpha}\right)=\delta_{\textbf{k}_{\alpha},\textbf{k}_{\beta}}$. The current is given by the expression:

\begin{equation}
\label{eq21}
\begin{split}
I&=\dfrac{8\pi e}{\hbar} \left(\dfrac{\hbar^2\kappa}{2mD}e^{-\kappa d}\right)^2 \sum_{B}\sum_{\textbf{k}_{\alpha},\textbf{k}_{\beta}}\lvert g_0 \left(\theta_L,\theta_R\right)\rvert^2\left[f_L\left(E_{\textbf{k},{\alpha}}\right)-f_R\left(E_{\textbf{k},{\beta}}\right)\right]\delta\left(e|V|-2\hbar v_Fk\right)\\
&=\dfrac{8\pi e}{\hbar} \left(\dfrac{\hbar^2\kappa}{2mD}e^{-\kappa d}\right)^2 \times 4u_{12}^4 \sum_{B,\textbf{k}}sin^2\theta_{\textbf{k}}\left[f_L\left(E_{\textbf{k},{\alpha}}\right)-f_R\left(E_{\textbf{k},{\beta}}\right)\right]\delta\left(e|V|-2\hbar v_Fk\right)\\
&=\dfrac{8\pi e}{\hbar} \left(\dfrac{\hbar^2\kappa}{2mD}e^{-\kappa d}\right)^2\dfrac{A}{2\pi}\times 2u_{12}^4 \int_{0}^{k_{max}} kdk\left[f_L\left(E_{\textbf{k},{\alpha}}\right)-f_R\left(E_{\textbf{k},{\beta}}\right)\right]\delta\left(e|V|-2\hbar v_Fk\right)\\
&=\dfrac{8\pi e}{\hbar} \left(\dfrac{\hbar^2\kappa}{2mD}e^{-\kappa d}\right)^2\dfrac{A}{\pi}\times u_{12}^4 \int_{0}^{e|V|/2\hbar v_F} kdk\left[f_L\left(E_{\textbf{k},{\alpha}}\right)-f_R\left(E_{\textbf{k},{\beta}}\right)\right]\delta\left(e|V|-2\hbar v_Fk\right)\\
&=\dfrac{8\pi e}{\hbar} \left(\dfrac{\hbar^2\kappa}{2mD}e^{-\kappa d}\right)^2\dfrac{A}{\pi}\times u_{12}^4 \times \dfrac{1}{2\hbar v_F}\times \dfrac{eV}{2\hbar v_F}\\
&=\dfrac{e^2A}{2\hbar}\left(\dfrac{\hbar \kappa u_{12}^2 e^{-\kappa d}}{mDv_F}\right)^2V
\end{split}
\end{equation}

\begin{figure}[H]
	\begin{center}
		\includegraphics[width=0.6\textwidth]{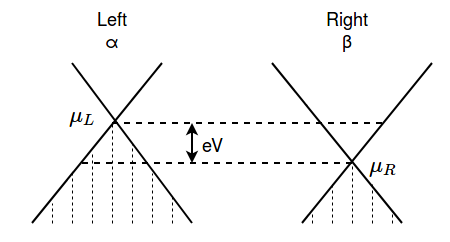}
		\mycaption[Band diagram of an undoped graphene]{Band diagram of an undoped graphene.}
		\label{fig:4a}
	\end{center}
\end{figure}

\noindent This is the expression for the current at zero temperature, when the graphene is undoped. it must be noted, that the current $I$ is a linear function of the voltage $V$. So, the $I-V$ characteristic will be a straight line. The band diagram for the undoped graphene sheet is given in Fig. \ref{fig:4a}.

Now, let us consider that the graphene is doped. For non resonant cases, it will be similar to undoped graphene, and the expression $\delta\left(E_{\textbf{k},{\alpha}}-E_{\textbf{k},{\beta}}\right)$ can be replaced by $\delta\left(eV-2\Delta E+\hbar v_Fk\right)$. Thus, the expression for current at zero temperature becomes

\begin{equation}
\label{eq22}
I=\dfrac{e^2A}{2\hbar}\left(\dfrac{\hbar \kappa u_{12}^2 e^{-\kappa d}}{mDv_F}\right)^2\left(\dfrac{2\Delta E}{e}-V\right)
\end{equation}

\noindent for $0<eV<2\Delta E$. Similarly, for $eV>2\Delta E$, the current is

\begin{equation}
\label{eq23}
I=\dfrac{e^2A}{2\hbar}\left(\dfrac{\hbar \kappa u_{12}^2 e^{-\kappa d}}{mDv_F}\right)^2\left(V-\dfrac{2\Delta E}{e}\right)
\end{equation}

\noindent Eq. \ref{eq22} and \ref{eq23} are for tunneling between like valleys. For tunneling between unlike valleys, $u_{12}$ will be replaced by $u_{11}$.

For $eV=2\Delta E$, i.e. the resonant case, $\delta\left(E_{\textbf{k},{\alpha}}-E_{\textbf{k},{\beta}}\right)=\delta\left(0\right)$. This expression being undefined, is a hindrance to calculation the expression for current seamlessly for the resonant case. However, we can calculate the expression for current for a finite graphene sheet with comparative ease, and it has been discussed in the next section.

\subsection{Tunneling current for finite graphene sheets}

By considering finite and symmetrical graphene sheets extending from $-L/2$ to $L/2$ along both the $x$ and $y$ axes and $A=L^2$, we get:

\begin{equation}
\label{eq24}
\begin{split}
\Lambda\left(\Delta\textbf{k}\right)&=\Big\lvert\dfrac{1}{A}\int_{-L/2}^{L/2}dx\int_{-L/2}^{L/2}dye^{i\Delta\textbf{k}\cdot\textbf{r}}\Big\rvert^2\\
&=\Big\lvert \text{sinc}\left(\dfrac{L\Delta k_x}{2}\right) \text{sinc}\left(\dfrac{L\Delta k_y}{2}\right) \Big \rvert^2
\end{split}
\end{equation}

\noindent where $\text{sinc}\left(x\right)\equiv \sin\left(x\right)/x$.The expression is maximum for $\textbf{k}_{\alpha}=\textbf{k}_{\beta}$. Substituting this value of $\Lambda\left(\Delta\textbf{k}\right)$ in the expression for the current does not allow us to evaluate the integral conveniently. So, we replace $\Lambda\left(\Delta\textbf{k}\right)$ with another function $\tilde{\Lambda}\left(\Delta\textbf{k}\right)$ which also has its maximum value at $\textbf{k}_{\alpha}=\textbf{k}_{\beta}$ and is given by:

\begin{equation}
\label{eq25}
\begin{split}
\tilde{\Lambda}\left(\Delta\textbf{k}\right)&=\exp\left(-\dfrac{A\lvert\Delta\textbf{k}\rvert^2}{4\pi}\right)\\
&=\exp\left(-\dfrac{A\Delta\textbf{k}_x^2}{4\pi}\right)\exp\left(-\dfrac{A\Delta\textbf{k}_y^2}{4\pi}\right)
\end{split}
\end{equation}

\noindent Now, by expressing $\lvert\Delta\textbf{k}\rvert^2=k_{\alpha}^2+k_{\beta}^2-2k_{\alpha}k_{\beta}\cos \theta$, where $\theta=\theta_L-\theta_R$, the angular part of the integral is expressed as:

\begin{equation}
\label{eq26}
\begin{split}
\int_{0}^{2\pi}d\theta_L\int_{0}^{2\pi}d\theta_R\lvert g_0 \left(\theta_L,\theta_R\right)\rvert^2\Lambda\left(\Delta\textbf{k}\right)&=\exp\left\lbrace-\dfrac{A}{4\pi}\left(k_{\alpha}^2+k_{\beta}^2\right)\right\rbrace\int_{0}^{2\pi}d\theta_L\\ &\int_{0}^{2\pi}d\theta_R\lvert g_0 \left(\theta_L,\theta_R\right)\rvert^2\exp\left(\dfrac{A}{2\pi}k_{\alpha}k_{\beta}\cos \theta\right)
\end{split}
\end{equation}

\noindent Now, substitute $\int_{0}^{2\pi}d\theta_L\int_{0}^{2\pi}d\theta_R\lvert g_0 \left(\theta_L,\theta_R\right)\rvert^2=8\pi^2\left[\left(u_{11}^4+u_{12}^4\right)I_0\left(Ak_{\alpha}k_{\beta}/2\pi\right)\pm u_{11}^4I_2\left(Ak_{\alpha}k_{\beta}/2\pi\right)\right]$ in Eq. \ref{eq26}, where $I_n$ is a modified Bessel function of first kind of order $n$. In the resonant case, $\delta\left(E_{\textbf{k},{\alpha}}-E_{\textbf{k},{\beta}}\right)=\delta\left(\hbar v_F k_{\alpha}-\hbar v_F k_{\beta}\right)=\delta\left(k_{\alpha}-k_{\beta}\right)/\hbar v_F$. Also, taking this into account, the expression for the current can be rewritten as:

\begin{equation}
\label{eq27}
\begin{split}
I&=\dfrac{32\pi e}{\hbar} \left(\dfrac{\hbar^2\kappa}{2mD}e^{-\kappa d}\right)^2 \dfrac{A^2}{\left(2\pi\right)^2\hbar v_F}\int k^2 dk \left[f_L\left(E_{\textbf{k},{\alpha}}\right)-f_R\left(E_{\textbf{k},{\beta}}\right)\right]\\
& \times \exp\left(-\dfrac{Ak^2}{2\pi}\right)\left[\left(u_{11}^4+u_{12}^4\right)I_0\left(\dfrac{Ak^2}{2\pi}\right)+ u_{11}^4I_2\left(\dfrac{Ak^2}{2\pi}\right)\right]
\end{split}
\end{equation}

\noindent Simplifying this expression and substituting the values of the Bessel functions we get the expression for the current in the resonant condition as:

\begin{equation}
\label{eq28}
I=\dfrac{0.8e^2A}{\sqrt{2\pi}\hbar}\left(\dfrac{\hbar\kappa e^{-\kappa d}}{mDv_F}\right)^2 \dfrac{L\Delta E^2\left(2u_{11}^4+u_{12}^4\right)}{e\hbar v_F}\exp\left\lbrace-\dfrac{A}{4\pi}\left[\dfrac{eV-2\Delta E}{\hbar v_F}\right]^2\right\rbrace
\end{equation}

\noindent The prominent Gaussian nature of the tunneling current is an interesting property to be noted.

%
%

\let\textcircled=\pgftextcircled
\chapter{Symmetric Tunneling Field Effect Transistor}
\label{chap:symfet}

\begin{figure}[]
	\begin{center}
		\includegraphics[width=0.45\textwidth]{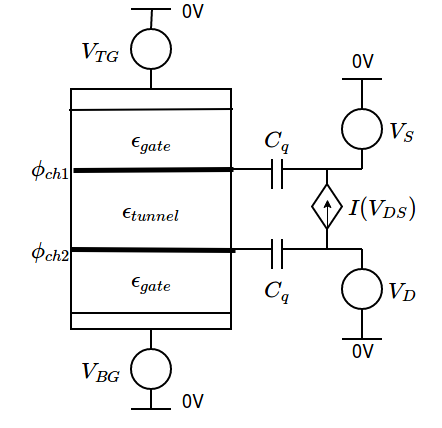}
		\mycaption[The SymFET device biasing circuit]{The SymFET device biasing circuit.}
		\label{fig:3a}
	\end{center}
\end{figure}

\initial{S}ymmetric Tunneling Field Effect Transistor (SymFET) is essentially a GIG junction FET as shown in Fig. \ref{fig:3a}. Two graphene layers sandwich a layer of an insulator. The gates are created on top of the graphene layers. Ohmic contacts are created with the graphene layers and the source (S) and drain (Drain) of the device. The graphene layer connected with the source is n-type and the graphene layer connected with the drain is p-type. The top and bottom gate voltages $V_{TG}$ and $V_{BG}$  modulate the Fermi occupancy level $\mu_n$ and $\mu_p$ in the top and bottom layers of graphene and are symmetric, i.e., $V_{TG}=-V_{BG}$.

\begin{figure}[]
	\begin{center}
		\includegraphics[width=0.7\textwidth]{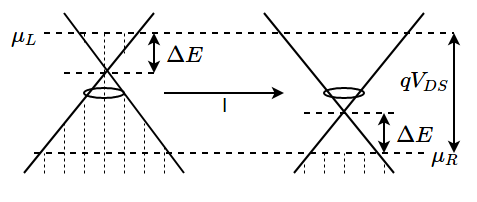}
		\mycaption[Band diagram for a doped GIG junction with $qV_{DS}>2\Delta E$]{Band diagram for a doped GIG junction with $qV_{DS}>2\Delta E$. Only a small current flows for the $\textbf{k}$-circle, midway between $\mu_L$ and $\mu_R$, for which energy and momentum conservation holds.}
		\label{fig:3b}
	\end{center}
\end{figure}

\begin{figure}[H]
	\begin{center}
		\includegraphics[width=0.45\textwidth]{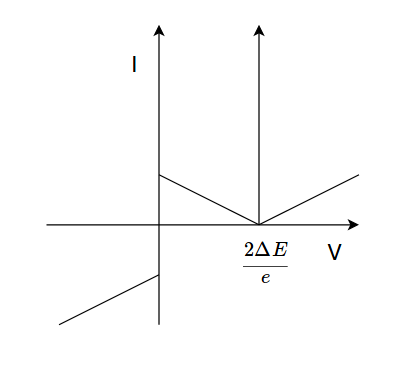}
		\mycaption[A qualitative $I-V$ characteristic of the SymFET]{A qualitative $I-V$ characteristic of the SymFET.}
		\label{fig:3c}
	\end{center}
\end{figure}

Under source-drain bias, when the n-graphene Dirac point is misaligned with the Dirac point of the p-graphene layer, energy and momentum conservation holds only for the value of the wavevector midway between the Fermi levels. Thus, only a small amount of current flows. In Fig. \ref{fig:3b} we can see the situation, when $qV_{DS}>2\Delta E$. Only a small current flows for the $\textbf{k}$-circle, midway between $\mu_L$ and $\mu_R$, for which energy and momentum conservation holds. From Eq. \ref{eq22} and \ref{eq23} we get the $I-V$ characteristics for $V<2\Delta E/e$ and $V>2\Delta E/e$, as shown in Fig. \ref{fig:3c}. When the Dirac points of both n and p type graphene are aligned, i.e., $V=2\Delta E/e$ energy and momentum values are conserved throughout the entire region between the two Fermi levels. Thus a large amount of current flows and the expression $\delta\left(E_{\textbf{k},{\alpha}}-E_{\textbf{k},{\beta}}\right)$ becomes $\delta\left(0\right)$ and blows up, as can be seen in Fig. \ref{fig:3c}.

A part of the applied voltage will fall across the graphene channels. The effect of finite density of states (DOS) is considered within the quantum capacitance of graphene. For the limiting case of temperature tending to $0$ K, we take the quantum capacitance as:

\begin{equation}
	C_q=\dfrac{2\lvert\Delta E\rvert}{\pi\left(\hbar v_F/q\right)^2}
\end{equation}

\noindent We can also use this for room temperatures, without much errors.

Let $\phi_{ch1}$ and $\phi_{ch2}$ be the channel potentials of the graphene layers. By applying charge neutrality. we get the following equations:

\begin{subequations}
	\begin{equation}
	\label{eq3a}
		\left(\frac{\phi_{ch1}}{q}+V_{TG}\right)C_g+	\left(\frac{\phi_{ch1}}{q}-\frac{\phi_{ch2}}{q}\right)C_t+\left(\frac{\phi_{ch1}}{q}-\frac{\mu_n}{q}\right)\frac{C_q}{2}+qN=0
	\end{equation}
	\begin{equation}
	\label{eq3b}
	    \left(\frac{\phi_{ch2}}{q}+V_{BG}\right)C_g+	\left(\frac{\phi_{ch2}}{q}-\frac{\phi_{ch1}}{q}\right)C_t+\left(\frac{\phi_{ch2}}{q}-\frac{\mu_p}{q}\right)\frac{C_q}{2}-qN=0
	\end{equation}
\end{subequations}

\noindent where the gate capacitance $C_g=\epsilon_g/t_g$, tunnel capacitance $C_t=\epsilon_t/t_t$ and $N=\Delta E_{\text{doping}}^2/\pi\left(\hbar v_F\right)^2$ is the doping concentration. We also assume the work functions of the metals match with undoped graphene for flatband condition at zero bias.

From, the band structure we can also write down the following relations:

\begin{subequations}
	\label{eq3c}
	\begin{equation}
	qV_{DS}=\mu_n-\mu_p
	\end{equation}
	\begin{equation}
	qV_{DS}=2\Delta E + \phi_{ch1} -  \phi_{ch2}
	\end{equation}
	\begin{equation}
	\Delta E=\mu_n- \phi_{ch1}= \phi_{ch2}-\mu_p
	\end{equation}
\end{subequations} 

\noindent by subtracting Eq. \ref{eq3b} from Eq. \ref{eq3a} and using relations in \ref{eq3c}, we get:

\begin{equation}
\left(V_{DS}-\frac{2\Delta E}{q}+2V_G\right)C_g +2\left(V_{DS}-\frac{2\Delta E}{q}\right)C_t-\frac{2q\Delta E^2}{\pi\left(\hbar v_F\right)^2}+2qN=0
\end{equation}

\noindent By solving this one gets 

\begin{equation}
\begin{split}
\Delta E\left(V_G,V_{DS}\right)&= -\dfrac{\left(2C_t+C_g\right)\pi\left(\hbar v_F /q\right)^2}{2} +
 \Big\lbrace \dfrac{\left(2C_t+C_g\right)^2\pi^2\left(\hbar v_F /q\right)^4}{4} \\
&+\dfrac{\pi\left(\hbar v_F\right)^2}{2q} \left[\left(V_{DS}+2V_G\right)C_g+2C_tV_{DS}+2qN\right] \Big\rbrace^{1/2}
\end{split}
\end{equation}

By substituting $G_1=\left(e^2A/2\hbar\right)\left(\hbar \kappa u_{12}^2 e^{-\kappa d}/mDv_F\right)^2$, we can rewrite Eq. \ref{eq22} and \ref{eq23} for non-resonant tunneling as:

\begin{equation}
	\label{eq3d}
	I=G_1\left(V_{DS}-\dfrac{2\Delta E}{q}\right)\text{sgn}\left(V_{DS}-\dfrac{2\Delta E}{q}\right)
\end{equation}

\noindent where $\text{sgn}\left(V_{DS}-2\Delta E/q\right)$ is $+1$ for $V_{DS}>2\Delta E/q$, $0$ for $V_{DS}=2\Delta E/q$ and $-1$ for $V_{DS}<2\Delta E/q$.

\noindent Similarly, the resonant tunneling current takes the form:

\begin{equation}
\label{eq3I}
	I=\dfrac{1.6}{\sqrt{2\pi}} G_1 \dfrac{L\Delta E^2\left(2u_{11}^4+u_{12}^4\right)}{u_{12}^4e\hbar v_F}\exp\left\lbrace-\dfrac{A}{4\pi}\left[\dfrac{eV_{DS}-2\Delta E}{\hbar v_F}\right]^2\right\rbrace
\end{equation}

Until now, we have only considered the current for zero temperature approximations. For finite temperatures, the expression from Eq. \ref{eq21} becomes

\begin{equation}
\begin{split}
I&=G_1\dfrac{4\hbar^2 v_F^2}{q} \int_{0}^{+\infty} k\left[ f\left(E_{n,\textbf{k}}-\mu_{n},T\right)-f\left(E_{p,\textbf{k}}-\mu_p,T\right)\right]\delta\left(2\Delta E-qV_{DS}-2\hbar v_Fk\right)dk \\
&=G_1\left(\dfrac{2\Delta E}{q}-V_{DS}\right)\left[ f\left(-qV_{DS}/2,T\right)-f\left(qV_{DS}/2,T\right)\right] \\
&=G_1\left(\dfrac{2\Delta E}{q}-V_{DS}\right)\tanh\left(\dfrac{qV_{DS}}{4k_BT}\right)
\end{split}
\end{equation}

\noindent for $qV_{DS}<2\Delta E$. Similarly, we can do this for $qV_{DS}>2\Delta E$. Thus, at finite temperature Eq. \ref{eq3d} becomes:

\begin{equation}
\label{eq3d}
I=G_1\left(V_{DS}-\dfrac{2\Delta E}{q}\right)\text{sgn}\left(V_{DS}-\dfrac{2\Delta E}{q}\right)\tanh\left(\dfrac{qV_{DS}}{4k_BT}\right)
\end{equation}

This gives us the finite temperature correction for non-resonant current. For the resonant case, i.e., when $qV_{DS}=2\Delta E$, we first note down the number of states for energy $E_d$ and temperature $T$

\begin{equation}
N_s\left(T\right)=\int_{-\infty}^{+\infty} \rho \left(E-E_d\right) \left[ f \left( E-E_d-\Delta E,T\right)-f\left(E-E_d+\Delta E,T\right)\right]dE
\end{equation}

\noindent where $\rho\left(E\right)=2\lvert E\rvert/\pi\left(\hbar v_F\right)^2$ is the DOS per unit area. With $N_s\left(0\right)=2\Delta E^2/\pi\left(\hbar v_F\right)^2$, we incorporate the effect of finite temperature in the resonant case, by multiplying Eq. \ref{eq28} with $N_s\left(T\right)/N_s\left(0\right)$. In terms of the Fermi-Dirac integrals $\mathscr{F}$ we have

\begin{equation}
\label{eq3e}
\dfrac{N_s\left(T\right)}{N_s\left(0\right)}=\dfrac{2\left(k_BT\right)^2}{\Delta E^2}\left[ \mathscr{F} \left(\dfrac{\Delta E}{k_BT}\right) - \mathscr{F} \left(\dfrac{\Delta E}{k_BT}\right) \right]
\end{equation}

\noindent where, 

\begin{equation}
\mathscr{F}\left(x\right)=\int_{0}^{\infty} \dfrac{t}{1+\exp\left(t-x\right)}dt
\end{equation}

Eq. \ref{eq22} and \ref{eq23} were derived for $A\rightarrow\infty$, and are valid for only large $L$. It is found that multiplying them with $\tanh\left(LqV_{DS}/\pi\hbar v_F\right)$ gives a satisfactory result. Again, in the product of Eq. \ref{eq3I} and \ref{eq3e}, we see that there is non zero current at $V_{DS}=0$ for samll values of $L$. This problem can be handled by multiplying the resonant current with the factor $\tanh\left(LqV_{DS}/2\pi\hbar v_F\right)$. Now, we can write the complete expression for the tunnel current as:

\begin{equation}
\begin{split}
	I&=G_1\left(V_{DS}-\dfrac{2\Delta E}{q}\right)\text{sgn}\left(V_{DS}-\dfrac{2\Delta E}{q}\right)\tanh\left(\dfrac{qV_{DS}}{4k_BT}\right)\tanh\left(\dfrac{LqV_{DS}}{\pi\hbar v_F}\right) \\
	&+ \dfrac{1.6}{\sqrt{2\pi}} G_1 \dfrac{L\Delta E^2\left(2u_{11}^4+u_{12}^4\right)}{u_{12}^4e\hbar v_F}\exp\left[-\dfrac{A}{4\pi}\left(\dfrac{eV-2\Delta E}{\hbar v_F}\right)^2\right] \times \dfrac{N_s\left(T\right)}{N_s\left(0\right)} \tanh\left(\dfrac{LqV_{DS}}{2\pi\hbar v_F}\right) 
\end{split}	
\end{equation}

\noindent Another thing that will be interesting to see, will be the ON-OFF ratio for the SymFET. If the peak current is $I_{ON}$ and the current at $V_{DS}\approx0$ is $I_{OFF}$ then

\begin{equation}
\label{eqonoff}
\dfrac{I_{ON}}{I_{OFF}} = \dfrac{0.8}{\sqrt{2\pi}}\dfrac{L\Delta E}{\hbar v_F}.
\end{equation}

We shall study how the $I-V$ characteristics and properties like the ON-OFF ratio change with different parameters of the device, in the next chapter.

%
%

\let\textcircled=\pgftextcircled
\chapter{SymFET Device Characteristics}
\label{chap:device}

\initial{T}he physics behind the SymFET device has been discussed in details and the current voltage relationships have been derived in the previous chapters. In the previous chapter, the final expression for the current was shown to be:

\begin{equation}
\begin{split}
I&=G_1\left(V_{DS}-\dfrac{2\Delta E}{q}\right)\text{sgn}\left(V_{DS}-\dfrac{2\Delta E}{q}\right)\tanh\left(\dfrac{qV_{DS}}{4k_BT}\right)\tanh\left(\dfrac{LqV_{DS}}{\pi\hbar v_F}\right) \\
&+ \dfrac{1.6}{\sqrt{2\pi}} G_1 \dfrac{L\Delta E^2\left(2u_{11}^4+u_{12}^4\right)}{u_{12}^4e\hbar v_F}\exp\left[-\dfrac{A}{4\pi}\left(\dfrac{eV-2\Delta E}{\hbar v_F}\right)^2\right] \times \dfrac{N_s\left(T\right)}{N_s\left(0\right)} \tanh\left(\dfrac{LqV_{DS}}{2\pi\hbar v_F}\right) 
\end{split}
\end{equation}

\noindent where the first expression comes from the treatment of the current at non-resonant conditions and the second term comes from the consideration of the resonant case. We have calculated the $I-V$ characteristics of the device for room temperature ($T=300$ K). We have set the decay constant $\kappa=17$ nm$^{-1}$ and the chemical doping level $\Delta E=0.1$ eV. The relative permittivity ($\epsilon_r$) of the insulator material, between the graphene layers has been taken to be $9$. The thickness of the gate is $t_g=20$ nm and the thickness of the tunneling portion is $t_t=0.5$ nm. The Fermi velocity $v_F$ is equal to $9\times10^5$ ms$^{-1}$. The coherent length of the device is set to be $L=100$ nm.

\section{Variation of current with drain-source voltage}

\begin{figure}[]
	\begin{center}
		\includegraphics[width=0.7\textwidth]{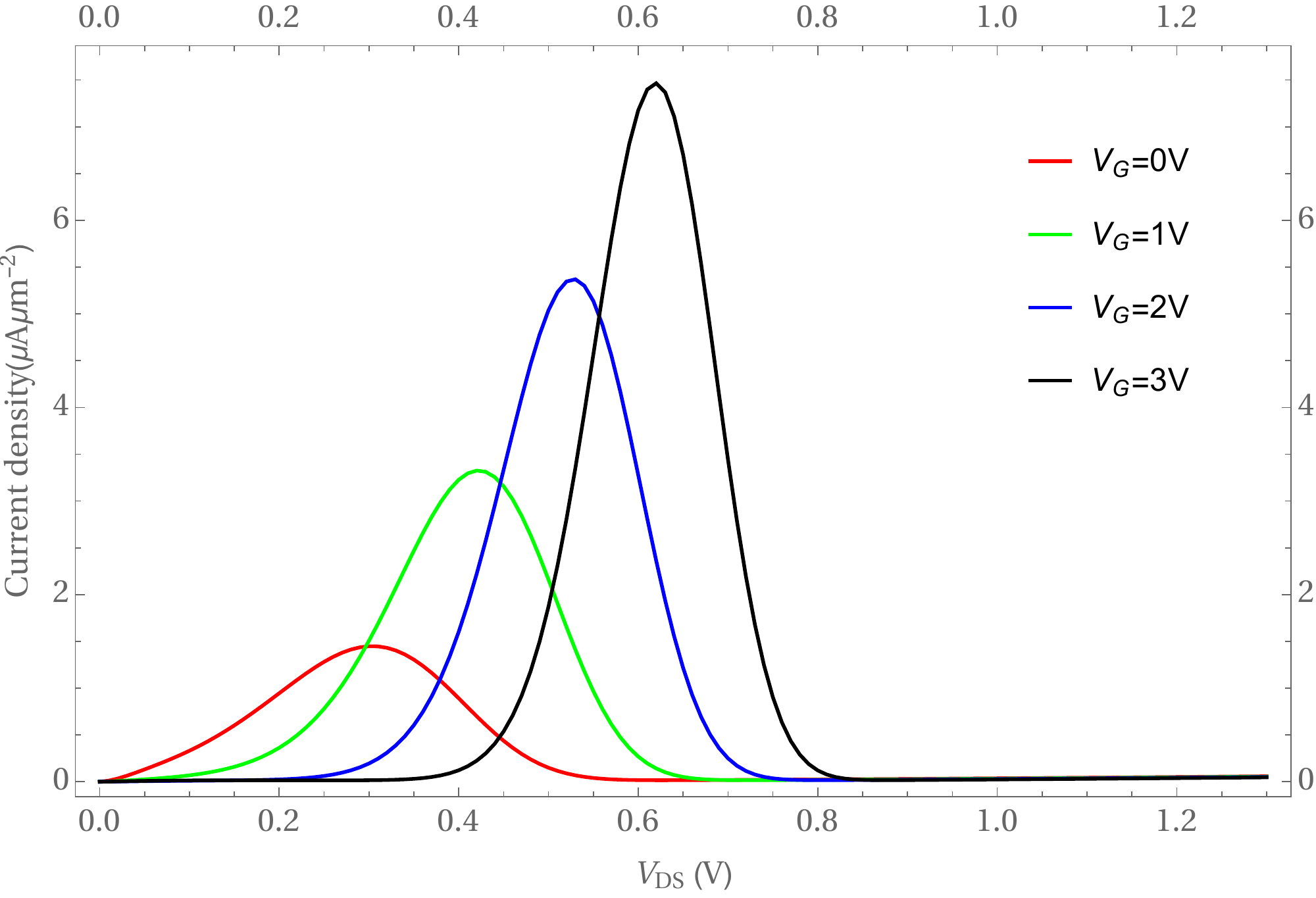}
		\mycaption[$I$ vs $V_{DS}$ characteristics for different values of $V_G$]{$I$ vs $V_{DS}$ characteristics for different values of $V_G$.}
		\label{ivds}
	\end{center}
\end{figure}

The $I-V$ characteristic of a SymFET with the afore mentioned parameters has been studied. It is important to note that, the current density of the device is a much more useful parameter to analyse its performance and characteristics and thus, the current has been converted to current density and expressed in the units \SI{}{\micro\ampere\micro\meter}$^{-1}$. The $I-V$ characteristic of the device is shown in Fig. \ref{ivds}. The variation of the characteristics with variation in the gate voltage is noted. An increase in the gate voltage increases the amount of tunneling current flowing through the device, and the resonant peak becomes sharper. Also, the value of the drain-source voltage $V_{DS}$ for which resonance occurs increases, with increase in gate voltage.

\begin{figure}[H]
	\begin{center}
		\includegraphics[width=0.6\textwidth]{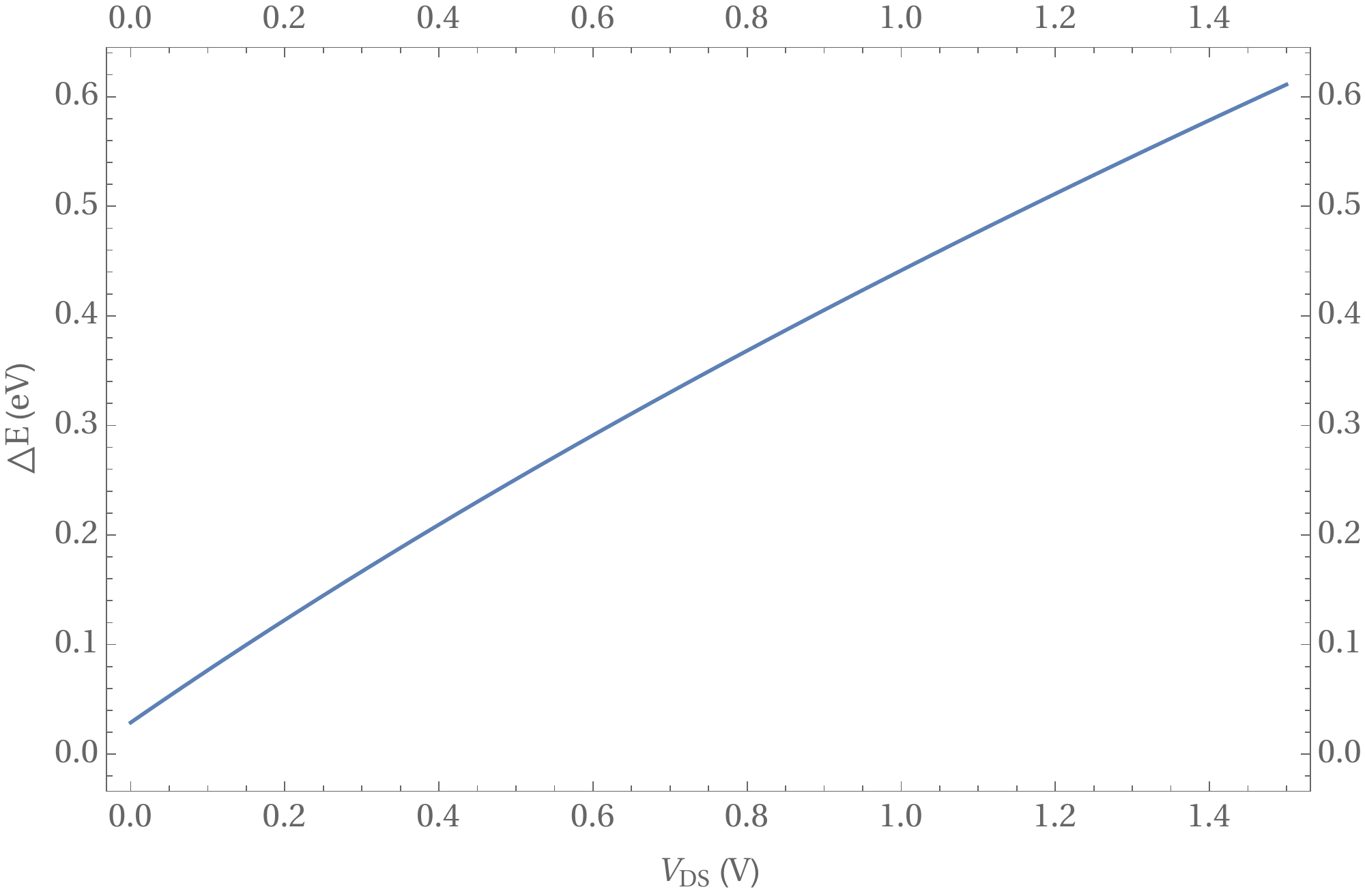}
		\mycaption[Dependence of $\Delta E$ on $V_{DS}$]{Dependence of $\Delta E$ on $V_{DS}$ at $V_G=2$ V.}
		\label{devds}
	\end{center}
\end{figure}

It is also interesting to note the variation of the Fermi level of graphene with applied drain-source voltage. The plot between $\Delta E$ and $V_{DS}$ for $V_G=2$ V is shown in Fig. \ref{devds}. The relation between $\Delta E$ and $V_{DS}$ is shown to be fairly linear in the range of voltage we are interested in.

\section{Variation of current with drain-source voltage}

\begin{figure}[]
	\begin{center}
		\includegraphics[width=0.7\textwidth]{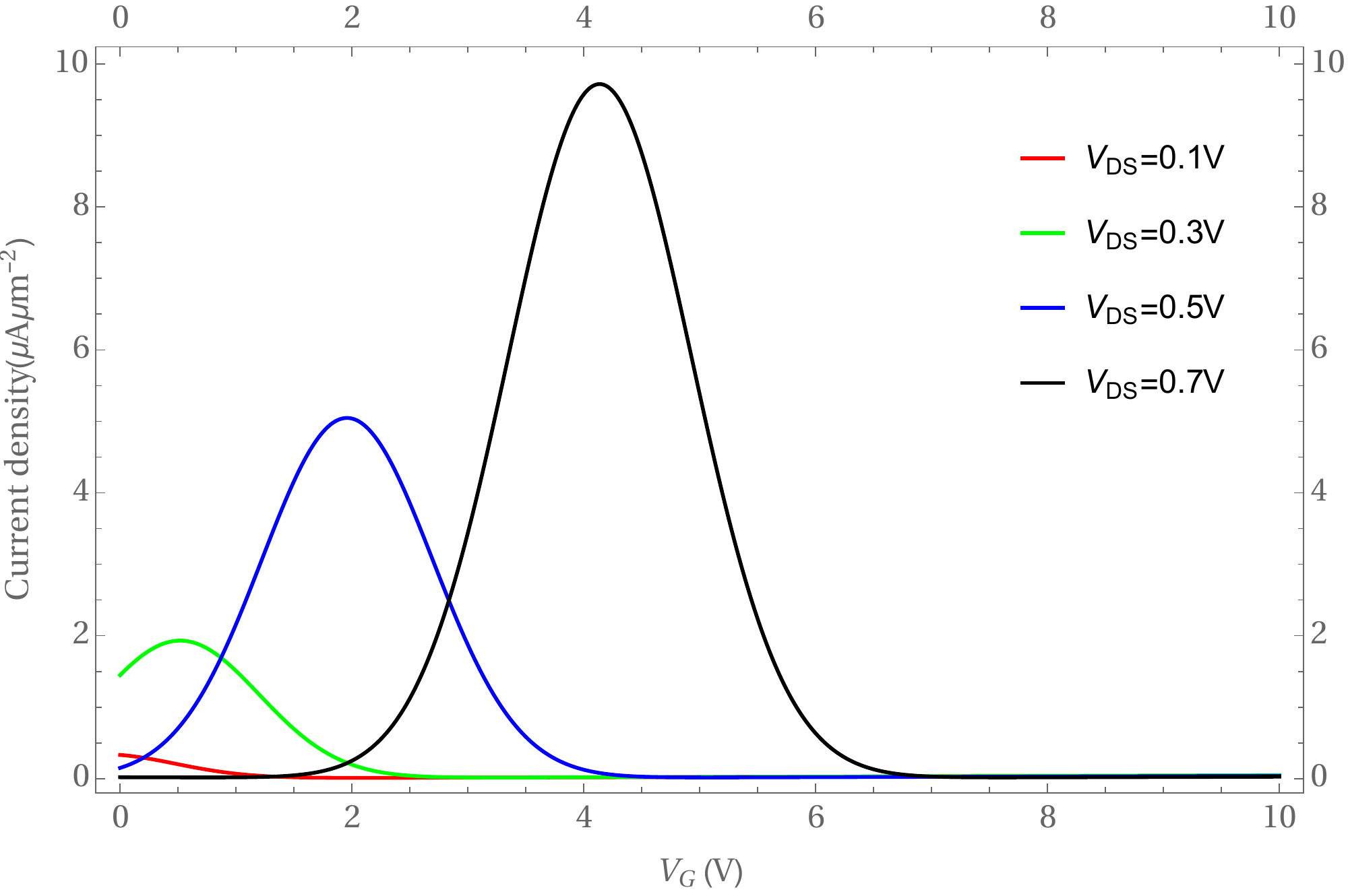}
		\mycaption[$I$ vs $V_{G}$ characteristics for different values of $V_{DS}$]{$I$ vs $V_{G}$ characteristics for different values of $V_{DS}$.}
		\label{ivg}
	\end{center}
\end{figure}

The transfer characteristic of the SymFET is shown in Fig. \ref{ivg}. As the drain-source voltage $V_{DS}$ is increased, the peak becomes higher, sharper and more prominent. Also the value of gate voltage for which the current peaks shifts to the right.

\section{Variation of device characteristics with device dimensions}

One of the most important properties of any electronic device is its physical dimension. We are always looking for higher degrees of circuit integration and so we are always interested in studying the performance of the device at lower dimensions. Other properties like leakage current and power dissipation are also direct functions of the dimensions of any device. Here, we will study how the $I-V$ characteristics of the SymFET change when its dimensions, particularly the coherence length $L$, the thickness of the tunneling insulator layer $t_t$ and the gate thickness $t_g$, are varied, and $V_G$ is constant at $3$ V.

\subsection{Variation in the coherent length}

\begin{figure}[H]
	\begin{center}
		\includegraphics[width=0.7\textwidth]{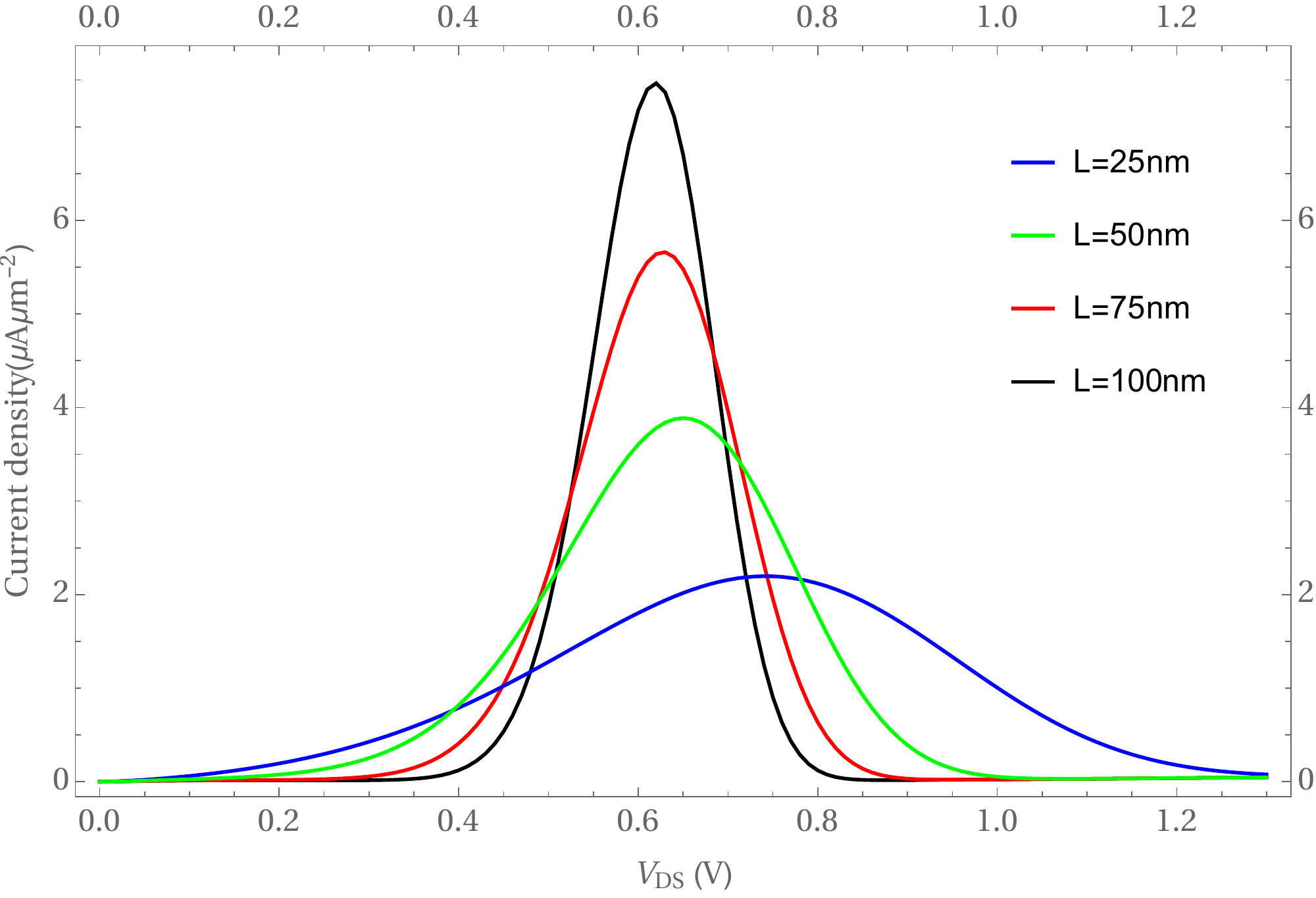}
		\mycaption[Variation in the $I-V$ characteristics with different coherent length $L$]{Variation in the $I-V$ characteristics with different coherent length $L$.}
		\label{varyl}
	\end{center}
\end{figure}

The coherent length $L$ (size of ordered area in graphene film) is the most important device dimensional parameter in our calculations. We have assumed $A=L^2$. Thus, the current density obtained from the device and properties like density of integration and heat dissipation are strong functions of the coherent length. The variation of the current-voltage relationships with varying coherent length has been shown in Fig. \ref{varyl}. Clearly the current capacity of the SymFET increases with an increase in coherent length. Also, the peaks become more well defined and symmetrical (a much wanted property) for higher values of $L$. 

\subsubsection{Dependence of on-off ratio on coherent length}

\begin{figure}[H]
	\begin{center}
		\includegraphics[width=0.5\textwidth]{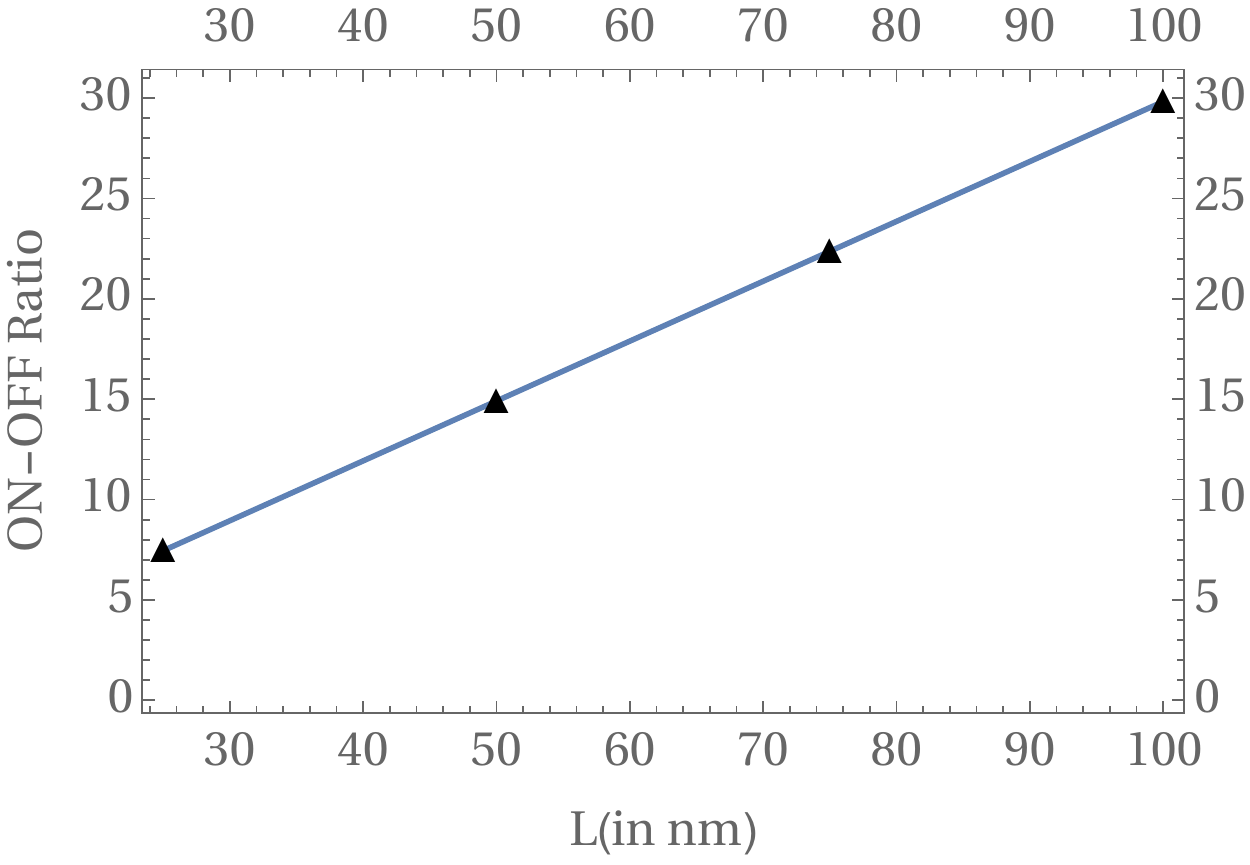}
		\mycaption[Variation in the ON-OFF ratio with change in coherent length $L$]{Variation in the ON-OFF ratio with change in coherent length $L$.}
		\label{onoff}
	\end{center}
\end{figure}

One of the biggest drawbacks of graphene devices, that have prevented their wide use in mainstream circuit manufacturing industry, is their much lower ON-OFF ratio compared to present day MOSFET devices. In Fig. \ref{onoff} we see the variation of the ON-OFF ratio as a function of the length. We can clearly see the linear dependence, as expected from Eq. \ref{eqonoff}. Thus, the deterioration of device performance in lower dimensions is evident.

\begin{figure}[H]
	\begin{center}
		\includegraphics[width=0.7\textwidth]{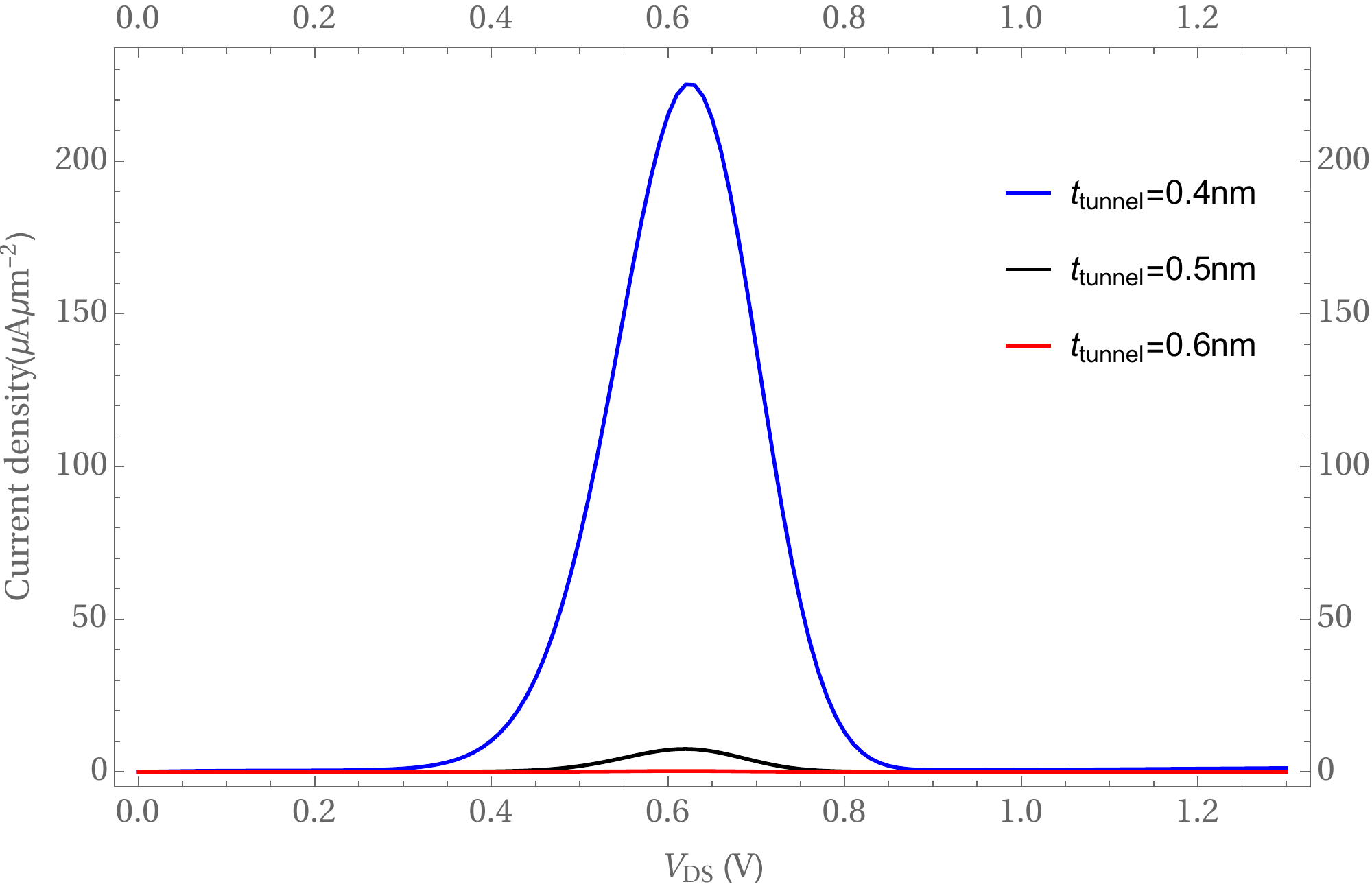}
		\mycaption[Variation in the $I-V$ characteristics with different $t_{tunnel}$]{Variation in the $I-V$ characteristics with different $t_{tunnel}$.}
		\label{ttunnel}
	\end{center}
\end{figure} 

\subsection{Variation in the thickness of tunneling insulator}

The thickness of the tunneling insulator is another important parameter governing the performance of the SymFET. Not, only does it contribute to the tunnel capacitance $C_t$, but also strongly contributes to the amount of tunneling current, due to the exponential decay of the wavefunction in the barrier, given by the expression $\exp\left(-\kappa d\right)$, where $\kappa$ is the decay constant in the barrier. In our calculations, $d=t_t$ and thus, even a small increase in the thickness of the insulator in the tunneling zone, causes a major drop in the resonant current peak. The variation of the tunneling current peak with changing thickness of the tunneling insulator is shown in Fig. \ref{ttunnel}.

\begin{figure}[]
	\begin{center}
		\includegraphics[width=0.7\textwidth]{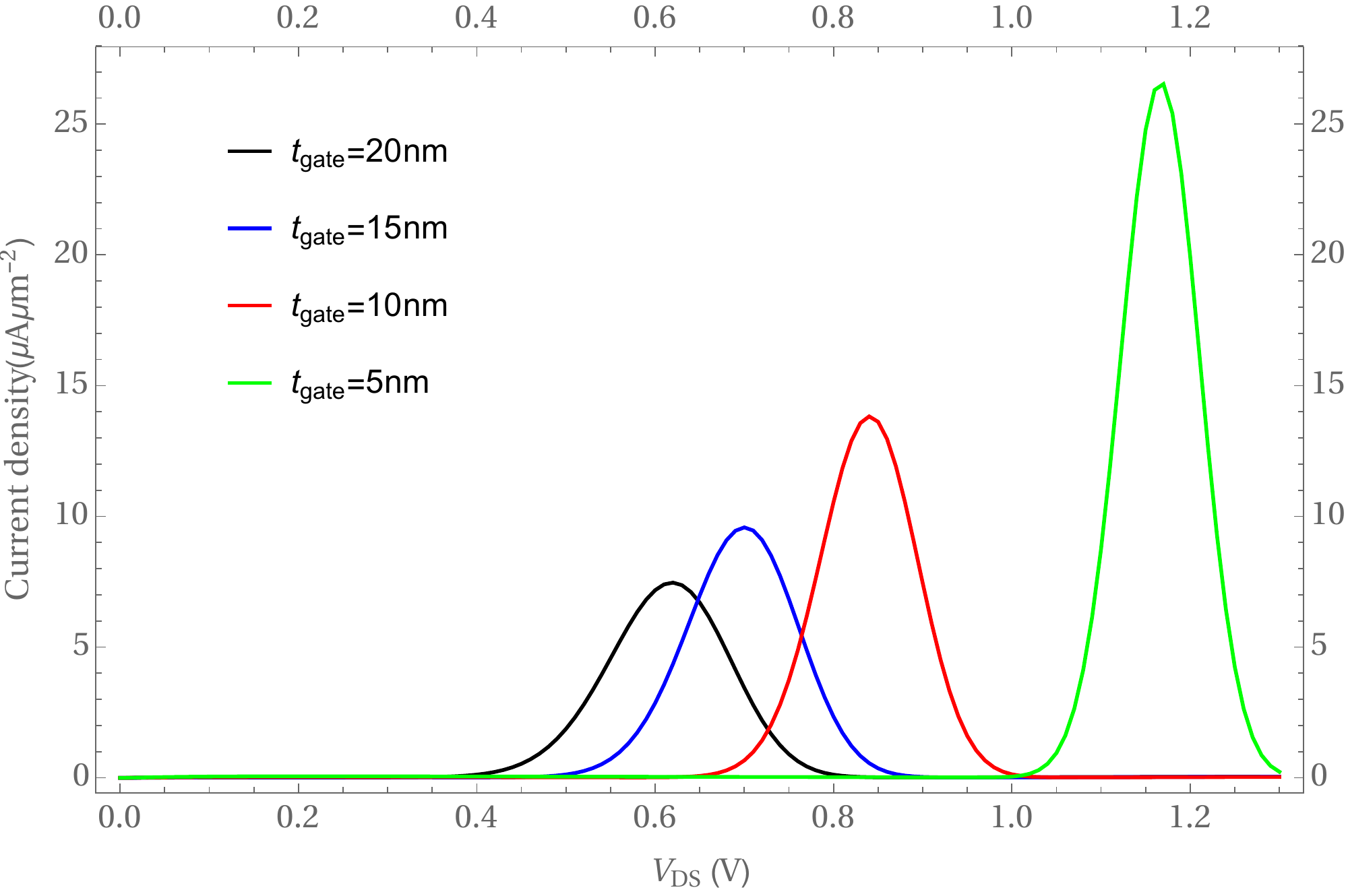}
		\mycaption[Variation in the $I-V$ characteristics with different $t_{gate}$]{Variation in the $I-V$ characteristics with different $t_{gate}$.}
		\label{tgate}
	\end{center}
\end{figure}

\subsection{Variation in the thickness of gate insulator}

Another important dimensional parameter of the SymFET is the gate thickness. The variation of the gate thickness changes the $I-V$ characteristics of the device, as shown in Fig. \ref{tgate}. As expected, the resonant current peak becomes higher and sharper with a decrease in the thickness of the gate insulator layer. Also, the value of $V_{DS}$ for which the current peaks increases with a decrease in $t_{gate}$.

\section{Variation of device characteristics with variation in temperature}

\begin{figure}[H]
	\begin{center}
		\includegraphics[width=0.7\textwidth]{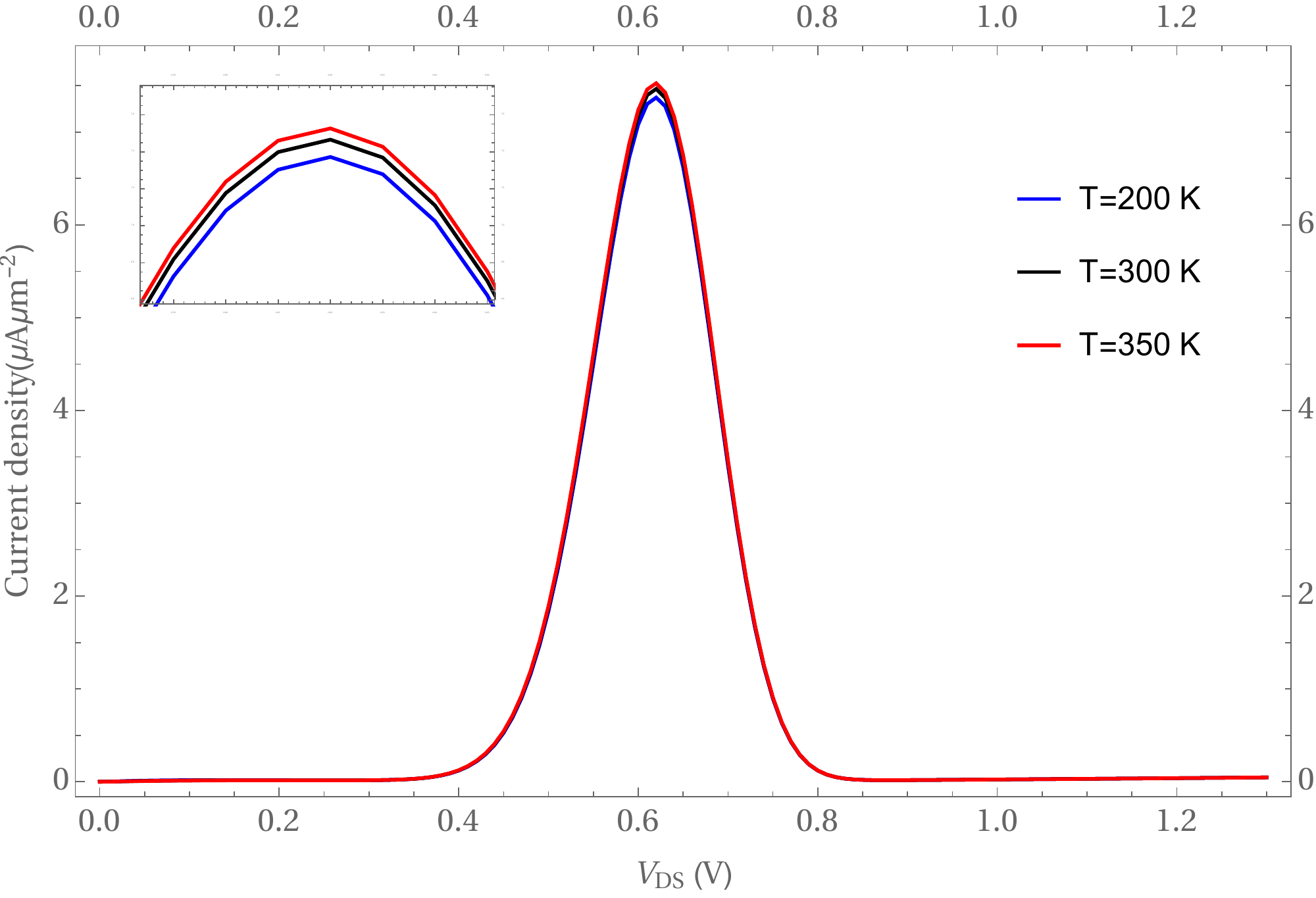}
		\mycaption[Variation in the $I-V$ characteristics with change in temperature]{Variation in the $I-V$ characteristics with change in temperature.}
		\label{temp}
	\end{center}
\end{figure}

The variation of the device characteristics with change in temperature is given in Fig. \ref{temp}. It can be seen that the SymFET is robust against temperature changes and there is negligible change or deterioration in the device performance due to temperature changes. The main reason behind the temperature insensitivity is that the charge transport is SymFET is governed by quantum tunneling.

\section{Variation of device characteristics with doping energy}

\begin{figure}[H]
	\begin{center}
		\includegraphics[width=0.7\textwidth]{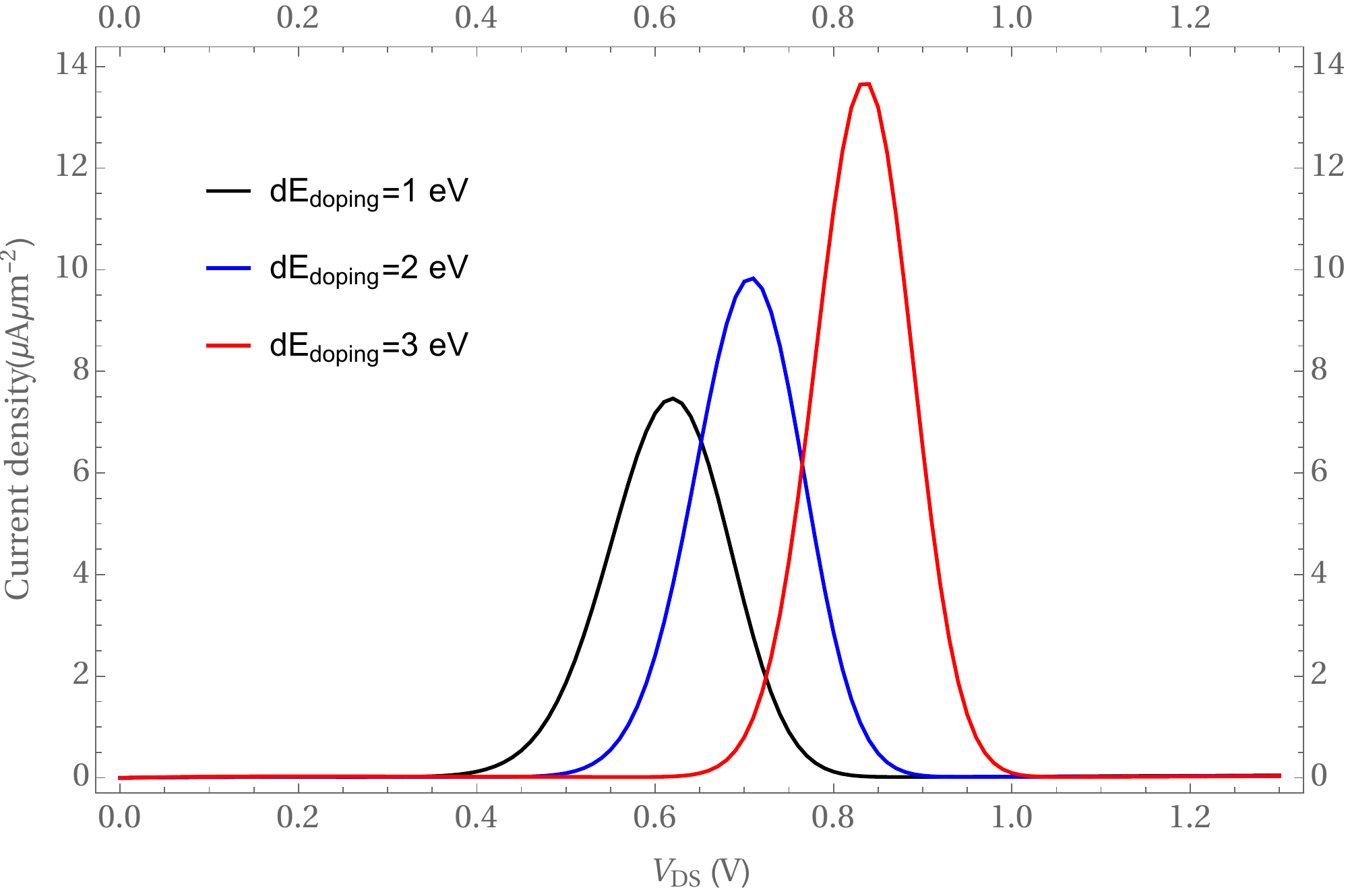}
		\mycaption[Variation in the $I-V$ characteristics with change in doping energy]{Variation in the $I-V$ characteristics with change in doping energy.}
		\label{doping}
	\end{center}
\end{figure}

Fig. \ref{doping} shows the variation in $I-V$ characteristics with doping energy. If the doping is increased, the resonant current peak increases. Also, the value of drain-source biasing at which the resonant current peaks increases.

\begin{figure}[H]
	\begin{center}
		\includegraphics[width=0.5\textwidth]{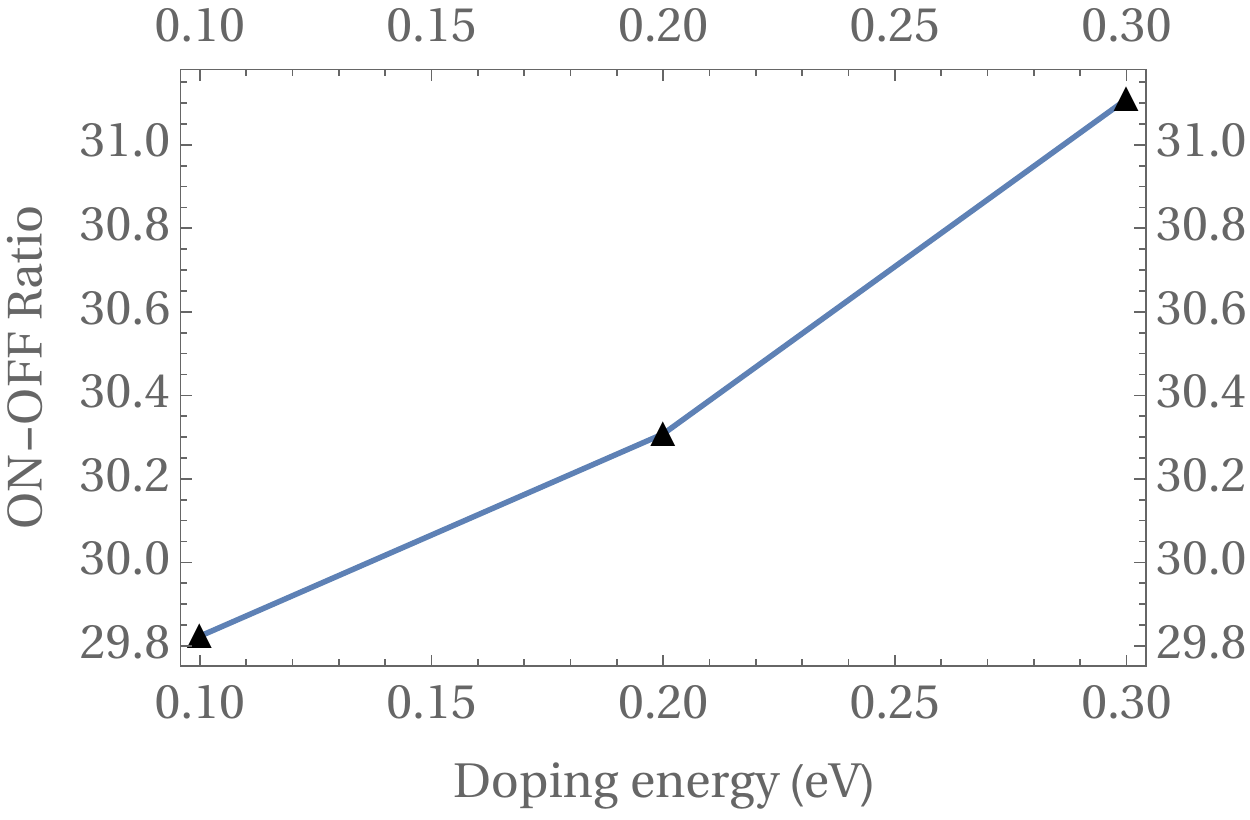}
		\mycaption[Variation in the ON-OFF ratio with change in doping energy]{Variation in the ON-OFF ratio with change in doping energy.}
		\label{onoffdoping}
	\end{center}
\end{figure}

Fig. \ref{onoffdoping} shows the variation in ON-OFF ratio with doping energy. The ON-OFF ratio of the device slightly increases if the doping levels are higher

\section{Variation of device characteristics with permittivity of dielectric}

\begin{figure}[H]
	\begin{center}
		\includegraphics[width=0.7\textwidth]{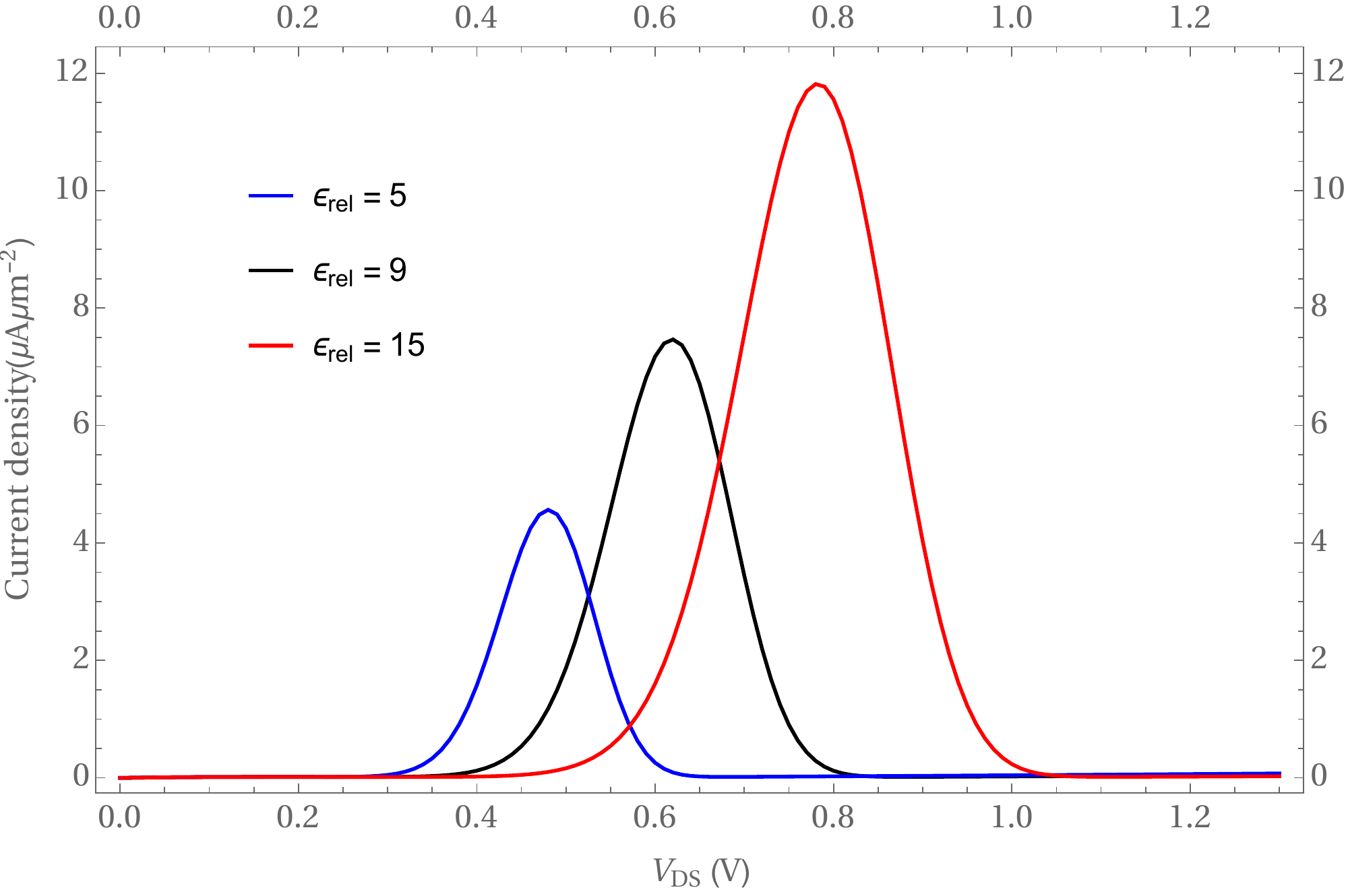}
		\mycaption[Variation in the $I-V$ characteristics with change in permittivity of insulator]{Variation in the $I-V$ characteristics with change in permittivity of insulator.}
		\label{ep}
	\end{center}
\end{figure}

Fig. \ref{ep} shows the variation in the $I-V$ characteristics of the SymFET with variation in permittivity of the dielectric material used. A greater resonant peak current is observed for insulators with greater permittivity.

%
%

\let\textcircled=\pgftextcircled
\chapter{Conclusion}
\label{chap:conclusion}

\initial{I}n this report, we briefly looked into the shortcomings of present day silicon metal oxide semiconductor field effect transistor or MOSFET devices and why graphene can be considered for the post-silicon era. Though many of the basic principles underlying the super properties of graphene have been know for about half a century, scientists started considering the use of two dimensional graphene sheets in electronic-device design, only in the past decade. In this report, we have shown the interlayer tunneling mechanism in a basic graphene-insulator-graphene (GIG) structure. Bardeen's transfer Hamiltonian approach has been used to derive the current-voltage relationships in such a device. The device allows only a small amount of tunneling current at most biasing voltages. But when the Dirac points of the to graphene layers are aligned, a large amount of current flows in the device, giving a symmetrical $I-V$ characteristic. 

Later, we use the formalism of tunneling in a GIG structure, to look into the working of a symmetric tunneling field effect transistor or SymFET. The SymFET proves to be very robust to temperature changes. The highly symmetrical resonant current peak, makes SymFET a good potential candidate for high speed analog and digital devices. The resonant current peak can be controlled by chemical doping and applying a gate bias. The resonant current peak increases with higher level of doping. If the gate bias is increased, the resonant peak becomes higher and sharper. Also, the value of the drain-source voltage for which the current peaks increases with increase in gate bias voltage. The characteristics of the device for varying device dimensions have been studied. Increasing the coherent length, increases the resonant current in the device, and the peak also becomes sharper. The tunneling current decreases exponentially with an increase in the thickness of the tunneling insulator, as one might expect. It happens due to the exponential attenuation of the wavefunction in the potential barrier between the graphene sheets. The tunneling current is also a strong function of the thickness of the gate insulator. The peak current falls with an increase in the thickness, but the decreases in peak current is much less pronounced in this case. Also, the value of drain-source voltage for which the tunneling current peaks decreases with an increase in the gate insulator thickness. From our investigations, e also see that, insulators with greater permittivity result in higher resonant peak current. Another interesting thing to note is that the ON-OFF ratio of the device. Decreasing the coherent length decreases the ON-OFF ratio, in a linear fashion. However, the ON-OFF ratio slightly increases, if the doping levels are increased. One of the main drawbacks of the SymFET, when compared to the MOSFET, is its small ON-OFF ratio.

Due to its ultrahigh mobility and symmetric bandstructures, such devices have many potential applications. The resonant peak behavior can be exploited to develop digital switching devices with much lower power consumption. We believe that we have only started to scratch at the surface of possibilities provided by graphene electronic devices. The field is still in its nascent phase and graphene may become the "supermaterial" to replace silicon in the future.
\appendix
\chapter{Appendix A}
\label{app:app01}

\initial{I}n this section, we shall derive Eq. \ref{eq8} given in Chapter \ref{chap:formalism}, using Bardeen's transfer Hamiltonian approach. Let us consider the system shown in Fig. \ref{fig:1}, where the barrier extends from $x_a$ to $x_b$ along the $x$ axis. There is metal $a$ to the left of $x_a$ and metal $b$ to the right of $x_b$. We can consider two many-particle states of the entire metal-barrier-metal system: $\psi_0$ and $\psi_{mn}$. $\psi_{mn}$ differs from $\psi_0$ in the transfer of an electron from state $m$ in metal $a$ to state $n$ in metal $b$. This leaves a hole in $m$ in $\psi_{mn}$. We must note that, the states $\psi_0$ and $\psi_{mn}$ can be specified by their quasi-particle occupation numbers in $a$ and $b$.

Let $\psi_0$ be the solution of the Schr\"odinger equation with energy $W_0$ to the left of $x=x_b$. Similarly, let $\psi_{mn}$ with energy $W_{mn}$ be the solution to the left of $x_a$, where wavefunction $\psi_n$ smoothly drops to zero. However, in the barrier both $\psi_0$ and $\psi_{mn}$ are applicable. So, the time dependent solution can be expressed as the linear combination of both the wavefunctions $\psi_0$ and $\psi_{mn}$:

\begin{equation}
\label{eq3a}
\psi=a\left(t\right)\psi_0e^{-iW_0t}+\sum_{m,n}b_{mn}\left(t\right)\psi_{mn}e^{-iW_{mn}t}
\end{equation}  

\noindent We have already seen upto this in Chap. \ref{chap:formalism}. 

With this in mind, we shall proceed to derive the expression for the tunneling current. The Schr\"odinger equation in region $a$ is given by:

\begin{equation}
H_a\psi=-\dfrac{\hbar^2}{2m}\nabla^2\psi\left(r\right)+V_a\left(r\right)\psi\left(r\right)
\end{equation}

\noindent and in region $b$ is:

\begin{equation}
H_b\psi=-\dfrac{\hbar^2}{2m}\nabla^2\psi\left(r\right)+V_b\left(r\right)\psi\left(r\right)
\end{equation}

\noindent Now, we can write the general wavefunction in the form:

\begin{equation}
\label{ansatz}
\psi=\psi e^{-i\varepsilon t}+\sum_{k}a_{k}\left(t\right)\phi_k
\end{equation} 

\noindent where $H_a\psi=\varepsilon\psi$ and $\phi_k$ are the bound states of $H_b$, with $H_b\phi_k=E_k\phi_k$. We know that $a_k\left(0-\right)=0$, i.e., $a_k\left(t\right)$ is zero just before tunneling occurs, and we need to approximate $a_k\left(t\right)$ for $t>0$.

Next, if we put Eq. \ref{ansatz} in the Schr\"odinger equation, we get (in units where $\hbar=1$):

\begin{equation}
\label{i}
\begin{split}
i \dfrac{\partial\psi\left(r\right)}{\partial t}&=H\left(\psi e^{-i\varepsilon t}\right)+\sum_{k}a_{k}\left(t\right)H\phi_k \\
&= e^{-i\varepsilon t} \left(H_a+\left(H-H_a\right)\right)\psi+ 
\sum_k a_k\left(t\right)\left(H_b+\left(H-H_b\right)\right)\phi_k\\
&= e^{-i\varepsilon t} \varepsilon \psi +e^{-i\varepsilon t} \left(H-H_a\right) \psi+\sum_k a_k\left(t\right)\left(E_k\phi_k+\left(H-H_b\right)\phi_k\right)
\end{split}
\end{equation}

\noindent By differentiating $\psi\left(r\right)$ w.r.t. time, we also get:

\begin{equation}
	\label{ii}
	\dfrac{\partial \psi\left(r\right)}{\partial t}=-i\varepsilon e^{-i\varepsilon t} \psi +\sum_k \dfrac{d}{dt}a_k\left(t\right)\phi_k
\end{equation}

\noindent From Eq. \ref{i} and \ref{ii}, we get:

\begin{equation}
\begin{split}
i\sum_k \dfrac{d}{dt}a_k\left(t\right)\phi_k =& e^{-i\varepsilon t}\left(H-H_a\right)\psi + \sum_k a_k\left(t\right)\left(E_k\phi_k+\left(H-H_b\right)\phi_k\right)
\end{split}
\end{equation}

\noindent Using the orthogonality property of $\phi_k$ we get:

\begin{equation}
	i\dfrac{d}{dt}a_j\left(t\right) = e^{-i\varepsilon t} \bra {\phi_j} H-H_a  \ket\psi +E_ja_j\left(t\right) + \sum_k a_k\left(t\right) \bra {\phi_j} H-H_b  \ket{\phi_k}
\end{equation}

\noindent Assuming $a_k\left(t\right)$ remains very small for a little while, even after $t>0$, we get:

\begin{equation}
i\dfrac{d}{dt}a_j\left(t\right) = e^{-i\varepsilon t} \bra {\phi_j} H-H_a  \ket\psi +E_ja_j\left(t\right)
\end{equation}

\noindent With initial condition $a_j\left(0\right)=0$, we get the solution of the differential equation as:

\begin{equation}
\begin{split}
	&a_j\left(t\right)=\dfrac{e^{-i\varepsilon t}-e^{-iE_j t}}{\varepsilon - E_j} \bra {\phi_j} H-H_a  \ket\psi \\
	&\therefore \lvert a_j\left(t\right) \rvert ^2= \dfrac{4\sin ^2\left(\frac{E_j-\varepsilon}{2}t\right)}{\left(E_j-\varepsilon\right)^2} \lvert \bra {\phi_j} H-H_a  \ket\psi \rvert ^2
\end{split}
\end{equation}

\noindent Now, $\lvert \braket{\phi_j}{\psi\left(t\right)}\rvert^2$ are the transition probabilities and 

\begin{equation}
	\braket{\phi_j}{\psi\left(t\right)}=e^{-i\varepsilon t}\braket{\phi_j}{\psi}+a_j\left(t\right)
\end{equation}

\noindent If $\braket{\phi_j}{\psi}$ is very small relative to $a_j\left(t\right)$, we get:

\begin{subequations}
\begin{equation}
\text{Transition probability}\approx \lvert a_j\left(t\right) \rvert ^2=\dfrac{4\sin ^2\left(\frac{E_j-\varepsilon}{2}t\right)}{\left(E_j-\varepsilon\right)^2} \lvert \bra {\phi_j} H-H_a  \ket\psi \rvert ^2
\end{equation}
\begin{equation}
\label{iii}
\begin{split}
	\therefore \text{Total tunneling rate}&= \dfrac{d}{dt} \sum_k \lvert a_k\left(t\right) \rvert ^2 \\ &= 4\dfrac{d}{dt} \left[ \sum_k  \dfrac{\sin ^2\left(\frac{E_j-\varepsilon}{2}t\right)}{\left(E_j-\varepsilon\right)^2} \lvert \bra {\phi_k} H-H_a  \ket\psi \rvert ^2 \right]
\end{split}
\end{equation}
\end{subequations}

\noindent The sum in Eq. \ref{iii} can be approximated by Fermi's Golden Rule since there are many $\phi_k$ states in region $b$. But it is only applicable at times $t$ large enough that the DOS per unit energy is nearly constant. Thus, the sum on the RHS of Eq. \ref{iii} can be rewritten as:

\begin{equation}
	\label{iv}
	\sum_k P_t\left(E_k-\varepsilon\right)M^2\left(\phi_k,\psi\right)
\end{equation}

\noindent where $M^2\left(\phi_k,\psi\right)= \lvert \bra {\phi_k} H-H_a  \ket\psi \rvert ^2$ and $P_t\left(x\right)=\sin^2\left(xt/2\right)/x^2$.

The function $P_t\left(x\right)$ is positive and its integral w.r.t. $x$ is equal to $\pi t/2$. But the main contribution to this integral comes from the interval $-4\pi/t <x<4\pi/t$. When $t$ is large enough the energy interval $-4\pi/t <E<4\pi/t$ becomes very narrow and the energy levels $E_k$ in region $b$ appear to be distributed with a constant energy over the energy interval. Let, $\varphi_b\left(\varepsilon\right)$ be the DOS at $\varepsilon$, i.e., the number of states per unit energy near $\varepsilon$ in region $b$. Letting $N_{\varepsilon}$ be the number of energy states in $b$ with energies in the interval $-4\pi/t+\varepsilon <E_k<4\pi/t+\varepsilon$ and setting 

\begin{equation}
	M^2\left(\psi\right)=\dfrac{1}{N_{\varepsilon}}\sum_{k:\lvert E_k-\varepsilon\lvert<4\pi/t}M^2\left(\phi_k, \psi\right)
\end{equation} 

\noindent we approximate Eq. \ref{iv} as follows:

\begin{equation}
	\begin{split}
	\sum_k P_t\left(E_k-\varepsilon\right)M^2\left(\phi_k,\psi\right)&\approx \sum_{k:\lvert E_k-\varepsilon\lvert<4\pi/t} P_t\left(E_k-\varepsilon\right)M^2\left(\phi_k,\psi\right) \\
	&\approx M^2\left(\psi\right) \sum_{k:\lvert E_k-\varepsilon\lvert<4\pi/t} P_t\left(E_k-\varepsilon\right) N_{\varepsilon} \\
	&\approx M^2\left(\psi\right) \varphi_b\left(\varepsilon\right) \int_{-4\pi/t}^{4\pi/t}P_t\left(E\right)dE \\
	&\approx M^2\left(\psi\right) \varphi_b\left(\varepsilon\right) \int_{-\infty}^{+\infty}P_t\left(E\right)dE \\
	&\approx M^2\left(\psi\right) \varphi_b\left(\varepsilon\right) \dfrac{\pi t}{2}
	\end{split}
\end{equation}

\begin{equation}
	\label{v}
	\begin{split}
	\therefore \dfrac{d}{dt} \sum_k \lvert a_k\left(t\right) \rvert ^2 &\approx \dfrac{d}{dt} \left(2\pi t M^2\left(\psi\right) \varphi_b\left(\varepsilon\right)\right) \\
	&=2\pi M^2\left(\psi\right) \varphi_b\left(\varepsilon\right)
	\end{split}
\end{equation}

Formula Eq. \ref{v} would be the rate at which electrons in the state $\psi$, in region $a$, are transferred into states in region $b$, if all of those states were vacant and available to receive electrons. Due to Pauli's exclusion principle, the DOS $\varphi_b\left(\varepsilon\right)$ needs to be multiplied by the fraction of unoccupied states in region $b$ with energies near $\varepsilon$.

Therefore, tunneling rate from $\psi$ to states $\phi_k$ in region $b$ is:

\begin{equation}
	\label{vi}
	2\pi M^2\left(\psi\right) \varphi_b\left(\varepsilon\right) \left(1-f_b\left(\varepsilon'\right)\right), \qquad \text{for }  \lvert \varepsilon'-\varepsilon\rvert < 4\pi/t
\end{equation}

\noindent The rate at which an electron in some state in region $b$ transports itself into state $\psi$ is:

\begin{equation}
\label{vii}
2\pi M^2\left(\psi\right) \varphi_b\left(\varepsilon\right) f_b\left(\varepsilon'\right), \qquad \text{for }  \lvert \varepsilon'-\varepsilon\rvert < 4\pi/t
\end{equation}

To find the net current we need to know which state in region $a$ are occupied and which are vacant. Occupied states contribute to a current of electrons from region $a$ to region $b$ at a rate given in Eq. \ref{vi} and vacant states enable electrons to flow from region $b$ to region $a$ at a a rate given in Eq. \ref{vii}. 

The current from $a$ to $b$ is the net rate of electron flow from region $a$ to region $b$, multiplied by the charge of electron.

\begin{equation}
	\therefore I_{ab}=2\pi e \sum_n \left[f_a\left(\varepsilon_n\right)\left(1-f_b\left(\varepsilon_n'\right)\right)-\left(1-f_a\left(\varepsilon_n\right)\right)f_b\left(\varepsilon_n'\right)\right] M^2\left(\psi_n\right) \varphi_b\left(\varepsilon_n\right), \quad \text{for } \lvert \varepsilon'-\varepsilon\rvert < 4\pi/t
\end{equation}

\noindent where $H_a\psi_n=\varepsilon_n\psi_n$ and 

\begin{equation}
M^2\left(\psi_n\right) \varphi_b\left(\varepsilon_n\right)=\sum_{k:\lvert E_k-\varepsilon\lvert<4\pi/t} \lvert \bra {\phi_k} H-H_a  \ket{ \psi_n} \rvert ^2
\end{equation}

\noindent such that 

\begin{equation}
\bra {\phi_k} H-H_a  \ket{ \psi_n} = \int \left[V\left(r\right)-V_a\left(r\right)\right]\phi_k^*\left(r\right)\psi_n\left(r\right) dr
\end{equation}

We choose any smooth surface in the barrier region that separates $a$ and $b$. Let $\partial T$ denote this separation and $T$ denote the region consisting of all points on the same side of $\partial T$ as the region $b$. The operator $\left(H-H_b\right)$ is the zero operator on the $b$  side of the separation surface. Now, let us shift from the unit $\hbar=1$ to $\hbar=\hbar$. 

\begin{equation}
\label{yo1}
	\begin{split}
	\therefore 0 &= \int_T \psi_n\left(r\right) \left(H-H_b\right)\phi_j^*\left(r\right)dr \\ 
	&=-\dfrac{\hbar^2}{2m}\int_T \psi_n\left(r\right) \nabla^2 \phi_j^*\left(r\right) dr +\int_T V\left(r\right) \psi_n\left(r\right) \phi_j^*\left(r\right) dr -E_j \int_T \psi_n\left(r\right) \phi_j^*\left(r\right) dr
	\end{split}
\end{equation}

On the other hand, in the same side $\left(H-H_a\right)$ is a non-zero operator.

\begin{equation}
\label{yo2}
	\begin{split}
	\therefore \bra {\phi_j} H-H_a  \ket{ \psi_n} &=  \int_T \phi_j^*\left(r\right) \left(H-H_a\right) \psi_n\left(r\right) dr \\&= -\dfrac{\hbar^2}{2m}\int_T \phi_j^*\left(r\right)  \nabla^2 \psi_n\left(r\right)dr +\int_T V\left(r\right)  \phi_j^*\left(r\right) \psi_n\left(r\right)  dr - \varepsilon_n \int_T \phi_j^*\left(r\right) \psi_n\left(r\right)  dr
	\end{split}
\end{equation}

\noindent Adding Eq. \ref{yo1} and \ref{yo2}, we get:

\begin{equation}
\begin{split}
\therefore \bra {\phi_j} H-H_a  \ket{ \psi_n} &= \int_T \phi_j^*\left(r\right) \left(-\dfrac{\hbar^2}{2m}\nabla^2 \psi_n\left(r\right)-\varepsilon_n\psi_n\left(r\right)\right)dr \\
&- \int_T \psi_n\left(r\right) \left(-\dfrac{\hbar^2}{2m}\nabla^2 \phi_j^*\left(r\right)-E_j\phi_j^*\left(r\right)\right)dr
\end{split}
\end{equation}

To obtain the tunneling current by Fermi's Golden Rule, we only consider matrix elements for which $\varepsilon_n$ and $E_j$ are approximately equal.

\begin{equation}
	\begin{split}
	\therefore \bra {\phi_j} E-E_a  \ket{ \psi_n} &\approx -\dfrac{\hbar^2}{2m} \int_T \nabla \cdot \left[\phi_j^*\left(r\right)\nabla \psi_n\left(r\right) - \psi_n\left(r\right) \nabla \phi_j^*\left(r\right)\right]dr \\
	& \approx -\dfrac{\hbar^2}{2m} \int_{\partial T}  \left[\phi_j^*\left(r\right)\nabla \psi_n\left(r\right) - \psi_n\left(r\right) \nabla \phi_j^*\left(r\right)\right]d\hat{S}
	\end{split}
\end{equation}

Thus, the tunneling current from $a$ to $b$ turns out to be:

\begin{equation}
\begin{split}
I_{ab}&=\dfrac{2\pi e}{\hbar} \sum_n \sum_{k:\lvert E_k - \varepsilon_n \rvert < 2h/t} \left[f_a\left(\varepsilon_n\right)\left(1-f_b\left(E_k\right)\right)-\left(1-f_a\left(\varepsilon_n\right)\right)f_b\left(E_k\right)\right] \lvert \bra {\phi_k} H-H_a  \ket{ \psi_n} \rvert ^2 \\
&= \dfrac{2\pi e}{\hbar} \sum_n \sum_{k} \left[f_a\left(\varepsilon_n\right)\left(1-f_b\left(E_k\right)\right)-\left(1-f_a\left(\varepsilon_n\right)\right)f_b\left(E_k\right)\right] \lvert \bra {\phi_k} H-H_a  \ket{ \psi_n} \rvert ^2 \delta\left(\varepsilon_n-E_k\right)
\end{split}
\end{equation}

\noindent where

\begin{equation}
\bra {\phi_k} H-H_a  \ket{ \psi_n} = -\frac{\hbar^2}{2m} \int_{\partial T}  \left[\phi_k\nabla \psi_n^* - \psi_n^* \nabla \phi_k\right]d\hat{S}
\end{equation}

\noindent For valley degeneracy $g_V$ and spin degeneracy $g_S$, the net current will be:

\begin{equation}
	I_{ab}=g_Sg_Ve\sum_{n,k}\left[\dfrac{1}{\tau_{nk}}f_a\left(\varepsilon_n\right)\left(1-f_b\left(E_k\right)\right)- \dfrac{1}{\tau_{kn}}f_b\left(E_k\right)\left(1-f_a\left(\varepsilon_n\right)\right)\right]
\end{equation}

\noindent where $1/\tau_{nk}$ and $1/\tau_{kn}$ are the tunneling rates for electrons. $f_a$ and $f_b$ are the Fermi occupation factors for the regions $a$ and $b$, where $f_{a/b}\left(E\right)=\left\lbrace1+\exp\left[\left(E-\mu_{a/b}\right)/k_BT\right]\right\rbrace^{-1}$. The tunneling rates are given by:

	\begin{equation}
	\dfrac{1}{\tau_{nk}} = \dfrac{1}{\tau_{kn}} = \dfrac{2\pi}{\hbar} \lvert M_{nk} \rvert^2 \delta\left(E_k - \varepsilon_n\right)
	\end{equation}
	
\noindent where	
	
	\begin{equation}
	M_{nk}= \dfrac{\hbar^2}{2m} \int_{\partial T} \left(\psi_n^* \nabla \phi_k - \phi_k\nabla \psi_n^* \right)d\hat{S}
	\end{equation}

\clearemptydoublepage
%
\backmatter
\bibliographystyle{unsrt}
\refstepcounter{chapter}
\bibliography{references}
\nocite{*}
\printindex
\end{document}